\documentclass[10pt,a4paper,Bold]{article}

\usepackage[utf8]{inputenc}
\usepackage[T1]{fontenc}
\usepackage{graphicx}
\usepackage[font=small]{caption}
\usepackage{copyrightbox}
\usepackage{url}
\usepackage[position=top]{subfig}
\usepackage{multicol,caption}
\usepackage{mathtools, cuted}
\usepackage{lipsum, color}
\usepackage{graphicx}	
\usepackage{amsmath}	
\usepackage{amssymb}	
\usepackage{multicol}        
\usepackage{bm}		
\usepackage{pdflscape}	
\usepackage[T1]{fontenc}
\usepackage{float}
\usepackage{blindtext}
\usepackage{xcolor}
\usepackage{pgfplots}
\usepackage{times}   
\usepackage{amsmath,scalerel}
\usepackage[toc,page]{appendix}
\usepackage[colorlinks=true, urlcolor=blue, linkcolor= blue,citecolor= blue, pdfborder={0 0 0}]{hyperref}
\usepackage[none]{hyphenat}
\newenvironment{Figure}
  {\par\medskip\noindent\minipage{\linewidth}}
  {\endminipage\par\medskip}

\title{Second multicols Demo}
\author{Overleaf}

\usepackage{geometry}
\begin{document}
\title{A Study of thin relativistic viscose accretion disk around a distorted kerr black hole (DKB) }
\author{Olya Layeghi$^{1}$\thanks{\texttt{o.layeghi@khayyam.ac.ir}} , 
Jamshid Ghanbari$^{1}$\thanks{\texttt{j.ghanbari@khayyam.ac.ir}} , 
Mahboobe Moeen Moghaddas$^{2}$\thanks{\texttt{Dr.moeen@kub.ac.ir}}\\
$^{1}$\ Department of Physics,Khayyam University,Mashhad,Iran\\
$^{2}$\ Department of Sciences,Kosar University of Bojnord,Bojnord, Iran\\
}
\date{\today}
\maketitle

                                                        
\begin{abstract}

In this paper, we analyze a thin disk around the distorted Kerr black hole (DKB) within the framework of general relativity using an axisymmetric solution of the Einstein equations. We consider this accretion disk around the Kerr black hole in an external gravitational field up to the quadrupole moment and discuss the key aspects of black hole accretion disk theory.
Our findings indicate that the presence of a quadrupole moment significantly influences the radiation emitted from the accretion disk. While the location of the innermost stable circular orbit (ISCO) remains largely unchanged, the magnitude of the radiation flux, as well as the shape, orientation, and energy distribution of the accretion disk, are affected. The direction of distortion of the event horizon determines whether the disk becomes more oblate or prolate, impacting observed variations in maximum height, position, and temperature. Furthermore, the quadrupole moment alters the geometry of the black hole's spacetime, which can influence the efficiency of energy extraction from the black hole's spin—an important factor in powering emissions from accretion disks. We obtain dynamical quantities around a distorted rotating black hole disk. Additionally, we examine how rotation influences the dynamics of the DKB. We also  investigate the effects of varying the viscosity coefficient on the behavior of the DKB.
\\
\end{abstract}

\begin{multicols}{2}


\section{Introduction}\label{sec.Intro}
A huge part of the light we receive from the depths of the universe reaches us from systems in which accretion played a major role. The luminosity of the disk is obtained from the Eddington luminosity relation, where \(L_{Edd}=1.3\cdot10^{38} (M/M_0 )\)  erg⁄s. Most analyzes of accretion disk models assume a steady and axisymmetric mode of matter accretion into a black hole. In these models, all physical quantities depend on only two spatial coordinates: the radial distance from the center r and the vertical distance from the plane of equatorial symmetry z. Most of the studied models assume that the disk is not vertically thick and the thin disk. In thin disks, inside the matter distribution $z/r\ll 1$ \cite{abramowicz2013foundations}. In models of thin accretion disks $(\dot{M}\le M_{Edd})$ where $\dot{M}$=16$L_{Edd}$/$c^2$ and Eddington's critical mass \(M_{Edd}=16 L_{Edd}/c^2\), the first studies were done by Shakura and Sunyaev \cite{shakura1973black}. The black hole distorts the space around it, which warps images of stars lined up almost directed behind it.\\
The distortion of spacetime around a black hole is not always perfectly spherical. The presence of external mass distributions or non-spherical internal structure can introduce a quadrupole moment, a measure of the non-spherical nature of the gravitational field.
Bardeen, Press and Teukolsky in 1972 developed techniques for studying perturbations and relativistic processes around rotating black holes \cite{bardeen1972rotating}. The first formal model of relativistic thin accretion disks was introduced by  Novika and Torne in 1973 \cite{novikov1973astrophysics}. For the first time in 1965, Doroshkevich considered a Schwarzschild black hole in a quadrupole gravitational field with a certain horizon \cite{doroshkevich1966gravitational}. Geroch and Hartl then did further analytical work on the warped black hole in 1982. They are obtained all exact solutions of Einstein's equation that represent static, axisymmetric black holes distorted by an external matter distribution \cite{geroch1982distorted}. In 2001, Fairhurst, Krishnan and Marolf presented new solutions to the Einstein-Maxwell equations representing a class of charged distorted black holes. These solutions are static-axisymmetric and generalizations of the distorted black hole solutions \cite{fairhurst2001distorted}. They discussed thermodynamics of such distorted black holes like Stoytcho S Yazadjiev \cite{yazadjiev2001distorted}. Shoom studied the interior of distorted stationary rotating black holes on the example of a Kerr black hole distorted by external static and axisymmetric mass distribution \cite{shoom2015distorted} and studied geodesic motion around a distorted Schwarzschild black hole \cite{shoom2016geodesic}. Shohreh Abdolrahimi has investigated the properties of the ergo region and the location of the curvature singularities for the Kerr black hole distorted by the gravitational field of external sources. The particular cases of quadrupole and octupole distortion are studied in detail \cite{abdolrahimi2015properties}. Grover et al studied the local shadow of the Schwarzschild black hole with a quadrupole distortion and the influence of the external gravitational field on the photon dynamics \cite{grover2018multiple}. A quadrupole specifically refers to the second-order term in a multipole expansion, characterizing deviations from spherical symmetry in the mass-energy distribution. While the centre of mass is balanced and it measures asymmetries in the gravitational field. In essence, quadrupole moments are a mathematical concept used to describe non-uniform distributions in space. The quadrupole moment significantly impacts the geometry of the black hole, leading to a departure from spherical symmetry and affecting the paths of particles and light around the black hole. In the simulations and analytical models of accretion disks, the energy content is considered in such a way that its effect on the space-time geometry is negligible.\\ 
Faraji and Hackmann constructed the relativistic standard steady, optically thick, cold, and geometrically thin accretion disk around a distorted Schwarzschild black hole and studied the effects due to a distortion up to the quadrupole and compare the physical characteristics of this disk to the usual Schwarzschild case \cite{faraji2020thin}. The work by Tao Zhu et al. on the properties of the electromagnetic spectrum emitted from the accretion disk around a static spherically symmetric black hole in 4EGB (four-dimensional Einstein-Gauss-Bonnet Black Hole) gravity adds to the growing body of research on thin accretion disks around black holes in different spacetime geometries. 
The study of accretion disks around black holes in various spacetime backgrounds has garnered significant interest, as evidenced by numerous works exploring this phenomenon.  For instance, Liu et al. \cite{liu2021thin} and \cite{liu2022thin} investigated thin accretion disks, providing a comprehensive overview of the field. Zhang et al. \cite{zhang2022image} focused on the imaging of a Bonner black dihole surrounded by a thin accretion disk, examining the impact of different radiation models on the observed images.  Further research by Panotopoulos et al. \cite{panotopoulos2021accretion} and \cite{panotopoulos2022binary} investigated accretion disks and soft spectral components of binary X-ray sources in massive gravity. Bubuianu et al. \cite{bubuianu2021nonassociative} explored solutions and toy models for four-dimensional black holes with distortions representing nonassociative star deformations.\\
This paper focuses on the DKB within the framework of general relativity, employing an axisymmetric solution of the Einstein equations.  In section \ref{sec.DKB}, we consider the Kerr black hole in an external gravitational field up to the quadrupole, as constructed by Abdolrahimi \cite{abdolrahimi2015properties}.  Novikov and Thorne \cite{novikov1973astrophysics} provided a detailed analysis of the Kerr spacetime, while Abramowicz \cite{abramowicz2013foundations} delved into the key aspects of black hole accretion disk theory.
In section \ref{sec.Metric}, we delve into the metric assumptions and Conservation Equations. We leverage the work of \cite{faraji2020thin}, who constructed a standard relativistic thin disk around a distorted Schwarzschild black hole up to the quadrupole.  We then proceed to solve the conservation equations in the DKB metric.  As the mass of the black hole is comparable to that of the sun (\(M_{Black hole}= M_{Sun}\)), the effect of self-gravity is negligible.\\
In section \ref{sec.result and plot}, we initially focus on the variations in the quadrupole moment while keeping the viscosity coefficient constant for both rotating and non-rotating configurations. Following this, we investigate a range of viscosity coefficient values for both distorted and undistorted states, ensuring that the rotational parameter remained fixed. Next, we assign specific viscosity coefficient values to the undistorted and distorted Kerr black holes, making necessary adjustments to the rotational parameter. Through detailed graphical representations, we perform a comprehensive analysis of their behaviors and the implications of these modifications. In section \ref{sec.discus and conclud}, we explain how our findings contribute to a deeper understanding of the dynamics involved in black hole configurations and their physical characteristics.
The results of our analysis, focusing on steady-state properties,such as flux,temperature, and accretion efficiency, which are derived from relativistic hydrodynamics and radiative transfer than classical thermodynamics. They are presented in the form of graphs and detailed explanations.  We choose to study the DKB due to the significance of rotation in black holes and the accuracy of the model in representing the motion of the accretion disk and the central black hole. additionally, we hope that our findings will align with future observations.  
Further details and explanations of the calculations are provided in Appendix \ref{sec.APPEN A}.
\section{Distorted Kerr Black hole}\label{sec.DKB}
Distortions are found in binary systems where the black hole forms an accretion disk with its compantion. They can be investigated numerically and analytically. many works have been done to numerically investigate such systems and merge black holes \cite{centrella2010black}. Also, some black holes can be modeled using axisymmetric and stationary solutions. In these cases, precise solutions for distorted black holes are examined.
The Kerr metric describes the geometry of spacetime around a charge-free rotating axisymmetric black hole, which is an exact solution of Einstein's field equations of general relativity. The DKB metric is given in the following form \cite{abdolrahimi2015properties}, \cite{breton1997kerr}:

\begin{equation}
\begin{split}
  ds^2&=-e^{2U}\frac{A}{B}(dt-\omega d\phi)^2\\
  &+\frac{1}{(1-(\beta)^2)^2}Be^{-2U+2V} (\frac{dx^2}{x^2-1}+\frac{dy^2}{1-y^2})\\
  &+\frac{B}{A}e^{-2U}(x^2-1)(1-y^2)d\phi^2,
  \end{split}
\end{equation}
where the metric functions are expressed by\\
\begin{equation}
 A = (x^2-1)(1+ab)^2-(1-y^2)(b-a)^2.
 \end{equation}
 
\begin{equation}
 B = [x+1+(x-1)ab]^2+[(1+y)a+(1-y)b]^2.
\end{equation}

\begin{equation}
\begin{split}
  C&=(x^2-1)(1+ab)[b-a-y(a+b)]\\
  &+(1-y^2)(b-a)[1+ab+x(1-ab)].
  \end{split}
\end{equation}
It is in prolate spheroidal coordinates $(t,x,y,\phi)$ that a system of curvilinear coordinates is defined where two sets of coordinate surfaces are obtained by revolving the curves of the elliptic cylindrical coordinates. The third set of coordinates comprises of planes passing through this axis.\\
the metric function $\omega$ can be represented as\\
\begin{equation}
\omega=2e^{(-2U)}\frac{C}{A}-\frac{4\beta}{1-\beta^2}e^{(-2q)}.
\end{equation}
We obtain the following expressions for the metric functions\\
\begin{equation}
a=-\beta e^{[2q(x-y)(1+xy)]}.
\end{equation}
\begin{equation}
b=\beta e^{[2q(x+y)(1-xy)]}.
\end{equation}
\begin{equation}
U=\sum_{n=0}^\infty a_n R^n P_n (\frac{xy}{R}).
\end{equation}
\begin{equation}
R=\sqrt{x^2+y^2-1}.
\end{equation}
\begin{equation}
V=\sum_{n,k=1}^\infty \frac{nk}{n+k} a_n R^n P_n.
\end{equation}
\begin{equation}
x=\frac{r}{M}-1 , y=cos(\theta).
\end{equation}
We denote by \(P_n\) the Legendre polynomials, which depends on the argument $xy/R$ in all the expressions. The solution is represented in the prolate spheroidal coordinates, $m>0$, $\beta\in(0,1)$
and $a_n=q$ ,$n\in N$ are real constants. So:
\begin{equation}
U=-\frac{q}{2}(x^2+y^2)+\frac{q}{2}(3x^2y^2+1).
\end{equation}

\begin{equation}
\begin{split}
  V&=2qx(y^2-1)+\frac{1}{4}q^2[(x^2+y^2-1)(x^2+y^2\\
 &-10x^2y^2)-x^2-y^2+9x^4y^4+1).
  \end{split}
\end{equation}
The multipole moments must satisfy the following condition:
\begin{equation}
\sum_{n\gg 0} a_{2n+1}=0,
\end{equation}
where $t\in(-\infty,+\infty), x\in(1,+\infty), y\in[-1,1], \phi\in[0,2\pi]$.\\
The distortion functions can be expressed in terms of Legendre polynomials of the first kind. \\
In Appendix the calculations are explained. 
So we derive:
\begin{equation}
\mathbb{A}=\frac{B}{x^8}.
\end{equation}
\begin{equation}
\mathbb{B}=\frac{1}{Mx^3\Omega}.
\end{equation}
\begin{equation}
\mathbb{C}=\mathbb{B}^2 (E-\Omega L)^2.
\end{equation}
\begin{equation}
\mathbb{D}=\frac{1}{x^8}(Ae^{2U}+4a^2).
\end{equation}
\begin{equation}
\mathbb{E}=\mathbb{A}+3a^2(x^{-4}-2x^{-6}+a^2x^{-8}).
\end{equation}
\begin{equation}
\mathbb{F}=\frac{L\sqrt{\mathbb{C}}}{Mx}.
\end{equation}
\begin{equation}
\mathbb{G}=E\sqrt{\mathbb{C}}.
\end{equation}
\begin{equation}
\mathbb{L}=\frac{2r^2}{3M} \mathbb{B} (\mathbb{C})^{\frac{1}{2}}f.
\end{equation}

\begin{equation}
 \begin{split}
     f&=\frac{3}{2M}\frac{1}{x^2(x^3-3x+2a)} \\
       & (x-x_0-\frac{3}{2}a \ln\frac{x}{x_0}-\frac{3(x_1-a)^2}{x_1(x_1-x_2)(x_1-x_3)}\\
       & \ln (\frac{(x-x_1)}{(x_0-x_1)}-\frac{3(x_2-a)^2}{x_2(x_2-x_1)(x_2-x_3)} \ln(\frac{(x-x_2)}{(x_0-x_2)}\\
       & -\frac{3(x_3-a)^2}{x_3(x_3-x_1)(x_3-x_2)} \ln(\frac{(x-x_3)}{(x_0-x_3)}.
  \end{split}
\end{equation}

and for the spin parameter, we have:
\begin{equation}
a=-\frac{2\beta}{1+\beta^2} e^{(-2\sum_{n=1} a_{2n+1})}=-\frac{2\beta}{1+\beta^2} e^{(-2q)} .
\end{equation}

\section{Thin Disk}\label{sec.Metric}

Accretion disks are fundamental to a wide range of astrophysical phenomena, acting as critical zones for energy release and jet formation. While significant progress has been made in understanding accretion disks around standard spherical or rotating black holes, emerging research suggests that many black holes in the universe deviate from these idealized models. Distortions caused by factors such as angular momentum distribution, gravitational wave interactions, or other astrophysical influences can profoundly impact the dynamics of surrounding accretion disks. Given this complexity, it becomes essential to establish a set of assumptions that will guide our calculations and conditions related to the DKB and its associated accretion disk:\\*[3mm]
	- The disk is geometrically thin $(h=H/r\ll1)$ in which H is half thickness of the disk so a standard relativistic thin disk around a Kerr black hole is assumed \cite{riffert1995relativistic} . \\*[3mm]
	- It is cold which means 
\begin{equation}
kT\ll GMm/r
\end{equation}
\\
    - The disk is optically thick which means that the mean free path of photons is short.\\*[3mm]
	- We focused on the region beyond the horizon that exhibits both Axial and mass symmetry, It is rotating and this distortion maybe related to the outer parts of the accretion disk.\\*[3mm]
	- The disk is in the equatorial plane, means $u^\theta$ vanishes. \\*[3mm]
	- Radiation is vertical when it emanates directly from the source above, forming a perpendicular angle with the surface underneath.\\*[3mm]
	- The standard $\alpha$ viscosity model is considered $(S_r\phi=\alpha P)$ where $\alpha$ is a free parameter and p is total pressure \cite{faraji2020thin}. \\*[3mm]
	- The internal energy density, radial pressure gradient, magnetic pressure and convection are neglected.\\*[3mm]
	- The subcritical accretion rate is defined as $\dot{M} \le M_{Edd}=16 L_{Edd}⁄c^2 $.\\*[3mm]


\subsection{Conservation Equations}\label{sec.conserv}
Relativistic thin accretion disk models describe the behavior of matter in accretion disks around compact objects such as black holes and neutron stars. These models assume that the disk is thin compared to its radius and that the motion of matter can be described by three fundamental equations that determine its radial structure.\\
The relativistic form of these equations accounts for the effects of special relativity, including time dilation and length contraction at high speeds. The conservation laws for mass, momentum, and energy are expressed in terms of rest mass density, four-velocity, stress-energy tensor, total energy density, and energy flux vector.
 \cite{abramowicz2013foundations}.\\
The equations we derive:\\
\\
The first equation describes the behavior of fluid flow in a radial direction under relativistic conditions:
\begin{equation}
h_{\mu\sigma}(T^{\sigma\nu})_{;\nu}=0,
\end{equation}
where $h^{\mu\nu}=u^\mu u^\nu + g^{\mu\nu}$ (the projecion tensor), and\\
$T^{\sigma\nu}$ is the stress-energy tensor.\\
The second equation is particles conservation:
\begin{equation}
(\rho u^\mu )_{;\mu}=0.
\end{equation}
In which $u^\mu$ is the four velocity of the fluid and $\rho$ is the rest mass density.\\
The third one is the energy conservation equation which is given as:
\begin{equation}
u_\mu T^{\mu\nu}_{;\nu}=0.
\end{equation}
These equations are derived from the principles of special relativity, which take into account the effects of time dilation and length contraction at high speeds. They provide a more accurate description of the behavior of matter in accretion disks around compact objects than classical physics.\\
The stress-energy tensor is expressed below:
\begin{equation}
T^{\mu\nu}=h^{\mu\nu}-Pg^{\mu\nu}+q^\mu u^\nu+ q^\nu u^\mu+S^{\mu \nu},
\end{equation}
where h is the the enthalpy density and P is the total pressure. Also $q^\nu$ is refer to the transverse energy flux and $S^{\mu \nu}$ is the viscous stress energy tensor.\\
In relativistic form, in absence of bulk viscosity, the viscose stress energy tensor is given by  
$S^{\mu \nu}=-2\lambda\sigma^{\mu\nu}$\\
in which $\lambda$ shows the dynamical viscosity and $\sigma^{\mu\nu}$ is the shear tensor.\\
 \\
The component $\sigma_{r\phi}$ is the only non-null component in the shear tensor.
 \begin{equation}
\sigma_{r\phi}=\frac{1}{2}(u_{r;\beta}h_{\phi}^{\beta}+u_{\phi;\beta}h_r^\beta)-\frac{1}{3}h_{r\phi}u_{;\beta}^\beta.
\end{equation}
Continuity equations (25), (26) and (27) require additional simplifying assumptions to construct analytical models of accretion disks.
The original thin disk model provides a set of assumptions that transform the complete system of partial differential equations into an algebraic nonlinear system, allowing us to obtain local solutions analytically.\\
We are only interested in vertically integrated quantities between \(z=-H\) and \(z=+H\). For example, the surface 
density of the disc is defined as
\begin{equation}
\Sigma(r)=\int_{-H}^{H} \rho(r,z)\, dz=2\rho H,
\end{equation}
where H is the half thickness of the disk.
The mass accretion rate is defined as \(\dot{M}=-2\pi r\Sigma u^r\). In the study of accretion disks, the rate at which matter moves toward the center and contributes to the accumulation of mass can be described by an equation that depends on the mass accretion rate \(\dot{M}\).\\
To directly obtain the equations for vertically integrated quantities based on the thin case assumption from the continuity equations, one can simply ignore the z-dependence of all physical quantities except for pressure P and radiation flux \(q^z\), which can be assumed to be \cite{compere2017self} :\\
\begin{equation}
\begin{split}
   P(r,z) & =p(r)(1-\frac{z^2}{H^2}), \\
     & q^z(r,z)=F(r)\frac{z}{H} , (z\ll H).
\end{split}
\end{equation}
Here F(r) is the radiation flux emitted from either the upper or lower side of the disk as a result of heat flow in the vertical direction,thus considering the \(q^z\) component.
Additionally, similar to equation (30), the vertically integrated viscous stress W is derived as:
\begin{equation}
W(r)=\int_{-H}^{H} S_{\hat r \hat \phi}(r,z) dz=2S_{\hat r \hat \phi} H.
\end{equation}
The $\alpha$ viscosity prescription is given by:
\begin{equation}
W=2 \alpha P H.
\end{equation}
The total pressure P is defined as the sum of the radiation pressure and the gas pressure:
\begin{equation}
P=P^{(gas)}+P^{(radiation)}=\frac{\rho k T}{m_p}+ \frac{a}{3} T^4.
\end{equation}
In which k represents Boltzmann’s constant , $m_p$ is the rest mass of the proton, \(a\) is the radiation density constant, and
T denotes temperature.
From \cite{faraji2020thin}, based on the work of Abramowicz et al., The pressure equation in the vertical direction can be derived from the conservation of energy equation: 
\begin{equation}
\frac{P}{\rho}=\frac{1}{2} \frac{(HL)^2}{r^4}.
\end{equation}
From the combination of energy conservation (27),particle number conservation (26) and radial momentum (25):
\begin{equation}
-4\pi r \frac{(E-\Omega L)^2}{\Omega_{,r}}\frac{F}{\dot{M}}=\int_{r_0}^{r} (E-\Omega L) L_{,r} dr.
\end{equation}
The energy and angular momentum per unit mass of circular motion in the equatorial plane are represented by \(E=-u_t\) and \(L=u_\phi\), respectively, while the angular velocity is denoted by \(\Omega=u^\phi/u^t\) .
the energy transport law is as follows:\\
\begin{equation}
a T^4=\bar{\kappa} \Sigma F.
\end{equation}
Where $\kappa$ is the optical opacity of the disk,specifically referring to free–free (ff) absorption opacity and electron scattering opacity (es):
\begin{equation}
\bar{\kappa}=\bar{\kappa}_{ff}+\bar{\kappa}_{es}.
\end{equation}
\begin{equation}
 \begin{split}
    \bar{\kappa}_{ff}&=(0.64\times 10^{23} cm^2 g^{-1})(\frac{\rho}{g cm^{-3}} (\frac{T}{k})^{-7/2}), \\
      & \bar{\kappa}_{es}=0.40 cm^2 g^{-1}.
 \end{split}
\end{equation}
To find the answers for this particular model, one must solve equations (30)-(39), which govern the movement of the thin accretion disk. We used equation (1) as the metric for the distorted Kerr black hole solution, modified in the equatorial plane where \(\theta=\pi/2\) or \(y=0\) in the coordinates \cite{torres2002accretion}. Some of the equations have been expanded and listed in Appendix. Finally, we have determined the total energy, angular velocity and angular momentum follows: \\
The total energy:
\begin{equation}
E=\frac{E_1 (2\Omega \omega-1)}{\sqrt{E_1 (1- 4\Omega \omega -\omega^2) -\Omega^2+\frac{ (x^2-1) (1-y^2)}{E_1}}}.
\end{equation}
The angular momentum per unit mass:
\begin{equation}
 \begin{split}
      & \Omega=\frac{-(\frac{d}{dr}(2\omega E_1))}{-\omega^2 E_1+\frac{d}{dr}(\frac{ (x^2-1) (1-y^2)}{E_1})} \\
      & +\frac{{\sqrt{\frac{d}{dr}(-2\omega E_1)^2-\frac{d}{dr}(-\omega^2 E_1 +\frac{(x^2-1) (1-y^2)}{E_1})\frac{d}{dr}(-E_1)}}}{-\omega^2 E_1+\frac{d}{dr}(\frac{ (x^2-1) (1-y^2)}{E_1})}.
 \end{split}  
\end{equation}            
The angular velocity:
\begin{equation}
L=\frac{\Omega -\omega^2 E_1+(\frac{ (x^2-1) (1-y^2)}{E_1})+2\omega E_1}{\sqrt{E_1 (1- 4\Omega \omega -\omega^2) -\Omega^2+\frac{ (x^2-1) (1-y^2)}{E_1}}},
\end{equation}
which:
\begin{equation}
E_1=\frac{A}{B} e^{2U}.
\end{equation}


\section{Results}\label{sec.result and plot}
We work on the metric equations of curved space time and simplified them, ultimately discovering new algebraic equations as a result. The obtained results are depicted here, where we compare the plots of Kerr's distorted and undistorted spacetime.
Since the distorted solutions are only applicable in a limited region near the horizon, we focus solely on the inner portion of the ISCO disc within this spacetime. we use for slowly rotating Black hole of \cite{jefremov2015innermost}. In the Kerr metric, the location of the event horizon depends on the spin parameter $a$. In Boyer-Lindquist coordinates, the radius of the outer event horizon is given by $x=M+\sqrt{M^2-a^2}$ \cite{boyer1967maximal}. For a Schwarzschild black hole ($a=0$), the horizon is at $r=2M$, but in a distorted Kerr black hole, the horizon location varies with the spin \cite{abdolrahimi2009interior}.
The effectiveness of our exact solution relies heavily on the specific quadrupole selected for implementation. Shoom \cite{shoom2016geodesic} elucidates the crucial role that this choice plays in determining the accuracy of the resulting solution. The constants involved are as follows: (All constants are in the CGS system)
\begin{equation}
\alpha=0.1
\end{equation}
While \(\alpha\)\ is considered constant in all the figures, we also examine new changes to the standard \(\alpha\)\ viscosity model affect the figures. 

\begin{equation}
M=M_{\odot}\simeq 1.99\cdot 10^{33} g
\end{equation}
\begin{equation}
L_{Edd}= 1.2\cdot 10^{38}\cdot \frac{M}{M_{\odot}} \frac{erg}{s}.
\end{equation}
\begin{equation}
c=3\cdot 10^{10} \frac{cm}{s}
\end{equation}
\begin{equation}
\dot{M}=\frac{16\cdot L_{Edd}}{c^2}.
\end{equation}
\begin{equation}
\dot{m}=\frac{\dot{M}\cdot c^2}{L_{Edd}}.
\end{equation}
\begin{equation}
G=6.67 \cdot 10^{-8} \frac{cm^3}{g\cdot s^2}
\end{equation}
\begin{equation}
k=1.38\cdot 10^{-16} \frac{erg}{K}
\end{equation}
\begin{equation}
m_{p}=1.67\cdot 10^{-24} g
\end{equation}
Now, from \cite{novikov1973astrophysics}, we have written the equations for DKB:

\begin{equation}
F=7\times 10^{26} \dot m  m^{-1}  (r_*)^{-3} \mathbb{B}^{-1} \mathbb{C}^{-\frac{1}{2}} \mathbb{L}.
\end{equation}

\begin{equation}
\Sigma=5\times \alpha^{-1} \dot{m}^{-1} ({r_*})^{\frac{3}{2}} \mathbb{B}^3 \mathbb{E} \mathbb{C}^{\frac{1}{2}} \mathbb{A}^{-2} \mathbb{L}^{-1}.
\end{equation}

\begin{equation}
h=10^5 \dot{m} \mathbb{A}^2 \mathbb{C}^{\frac{1}{2}} \mathbb{L} ({r_*})^{-1} \mathbb{B}^{-3} \mathbb{D}^{-1} \mathbb{E}^{-1}.
\end{equation}

\begin{equation}
T=2\times10^7 \mathbb{B}^{\frac{1}{2}} \mathbb{E}^{\frac{1}{4}} \alpha^{-\frac{1}{4}} m^{-\frac{1}{4}} ({r_*})^{-\frac{3}{8}} \mathbb{A}^{-\frac{1}{2}}.
\end{equation}

\begin{equation}
P=5\times10^{-5} m^{\frac{7}{4}} \alpha^{-\frac{1}{4}} \dot{m}^{-2} ({r_*})^{\frac{21}{8}} \mathbb{B}^{\frac{9}{2}} \mathbb{D} \mathbb{E}^{\frac{5}{4}} \mathbb{A}^{-\frac{5}{2}} \mathbb{L}^{-2}.
\end{equation}

\begin{equation}
W=-7\times 10^{26} (\frac{d \Omega}{dr})^{-1}  \dot m  m^{-1}  (r_*)^{-3} \mathbb{B}^{-1} \mathbb{C}^{-\frac{1}{2}} \mathbb{L}.
\end{equation}
where $r_*\simeq4$. In the following sections, based on equations 53 to 58, we present graphs illustrating the flux, surface density, half-thickness, temperature, pressure, and viscosity as functions of the radial coordinates (distance from the black hole). To determine the event horizon of the DKB, we assumed $M=1$. These graphs account for various states specified in the captions of the figures:\\
\\

\begin{Figure}
 \centering
 \advance\leftskip-2cm
 \advance\rightskip-2cm
 \includegraphics[width=8cm, height=7cm]{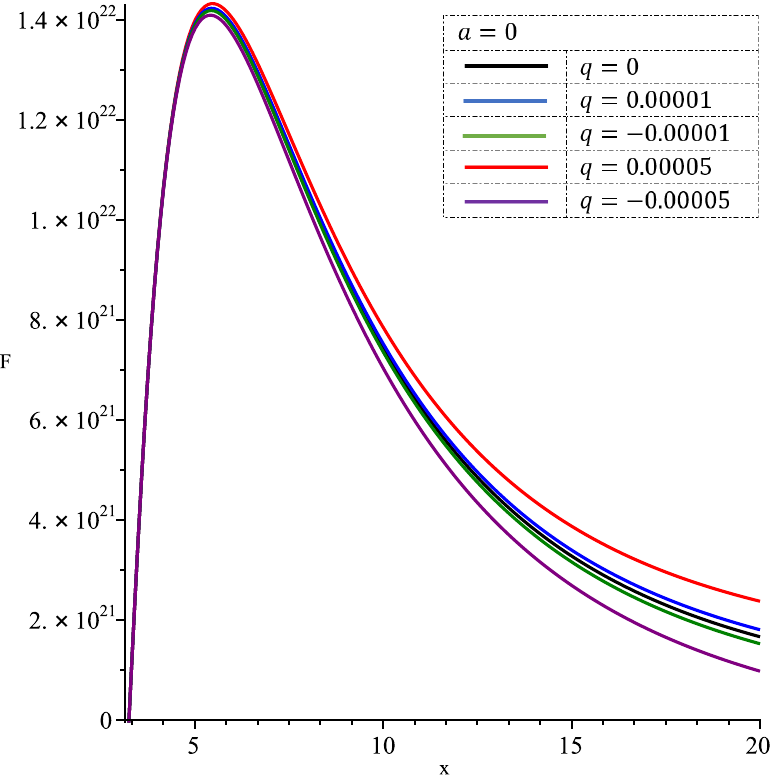}
 \captionof{figure}{Radiation flux F for \(q=0\) and \(q\neq0\) in the $(\frac{erg}{cm^2\cdot s})$ unit and \(a=0\).}
\end{Figure}

\begin{Figure}
 \centering
 \advance\leftskip-2cm
 \advance\rightskip-2cm
 \includegraphics[width=8cm, height=7cm]{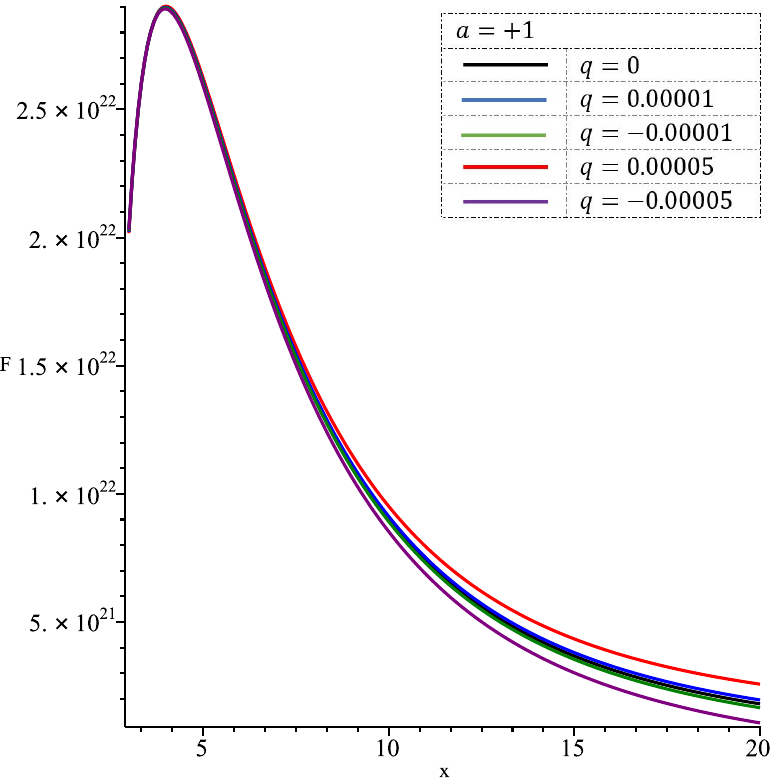}
 \captionof{figure}{Radiation flux F for \(q=0\) and \(q\neq0\) in the $(\frac{erg}{cm^2\cdot s})$ unit and \(a=+1\).}
\end{Figure}

Figures 1 and 2 suggest that the presence of a quadrupole moment in the black hole affects the radiation emitted from its accretion disk. The differences at the beginning of the graphs indicate that rotation causes the flux to increase more rapidly. Consequently, the magnitude of the radiation flux is influenced by the quadrupole moment: positive quadrupoles $(q=0.00001$ and $q=0.00005)$ result in higher radiation flux, while negative quadrupoles $(q=-0.00001$ and $q=-0.00005)$ lead to lower radiation flux.\\
As expected, when considering rotation, the maximum flux approximately doubles and exhibits a steeper increase in subsequent locations compared to the case without rotation.

\begin{Figure}
 \centering
 \advance\leftskip-2cm
 \advance\rightskip-2cm
 \includegraphics[width=8cm, height=7cm]{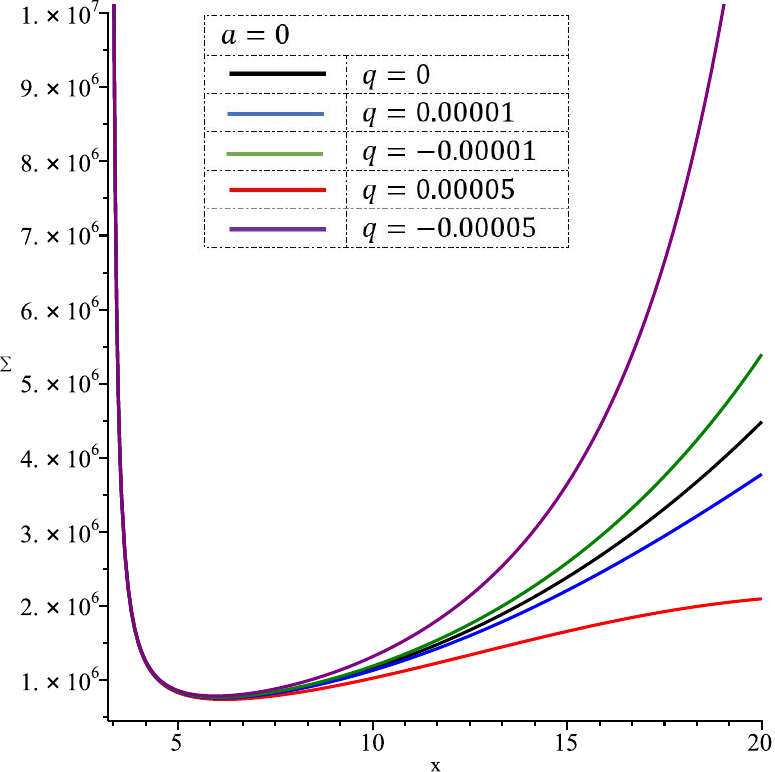}
 \captionof{figure}{Surface densiry ${\Sigma}$ for \(q=0\) and \(q\neq0\) in the $(\frac{g}{cm^2})$ unit and \(a=0\).}
\end{Figure}

\begin{Figure}
 \centering
 \advance\leftskip-2cm
 \advance\rightskip-2cm
 \includegraphics[width=8cm, height=7cm]{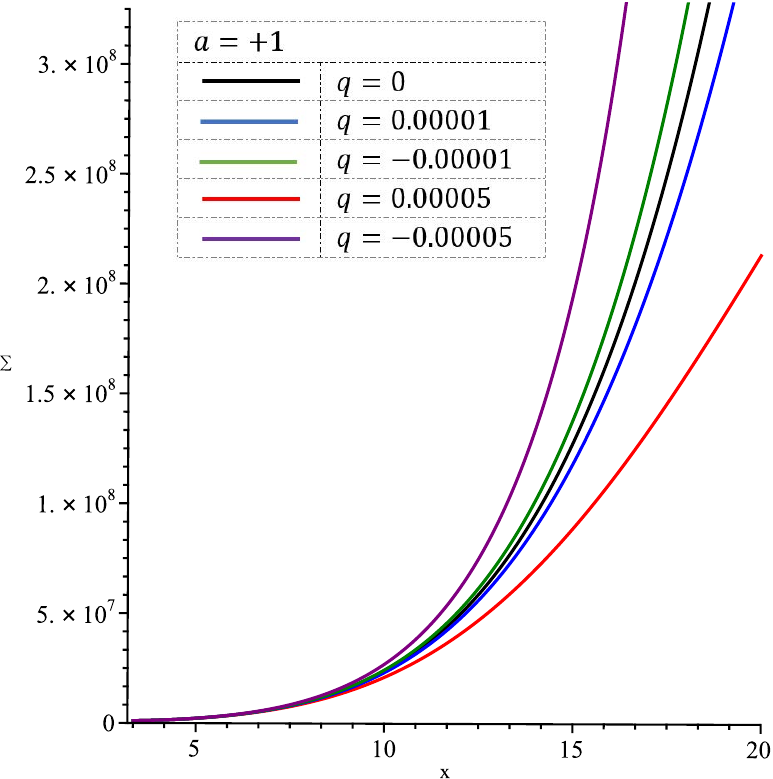}
 \captionof{figure}{Surface density ${\Sigma}$ for \(q=0\) and \(q\neq0\) in the $(\frac{g}{cm^2})$ unit and \(a=+1\).}
\end{Figure}

In figures 3 and 4, the radial coordinate x is plotted against the surface density ${\Sigma}$. At x = 3, we observe the maximum surface density without rotation at the initial point. However, when rotation is applied, this initial maximum shifts closer to the horizon. When we compare each quadrupole moment q in both states, we notice that in the compare rotation state (Kerr), the growth of surface density occurs more rapidly. This indicates the surface density decreases with the distance from the horizon, which we expect, but by adding rotation It is increasing that needs to be investigated more so that in higher positions it should tend to zero in order to get the correct result.\\
The surface density plots for positive $(q=0.00001$ and $q=0.00005)$ quadrupoles increase slowly, while those for negative $(q=-0.00001$ and $q=-0.00005)$ quadrupoles increase rapidly, both in comparison to the undistorted Kerr spacetime.\\

\begin{Figure}
 \centering
 \advance\leftskip-2cm
 \advance\rightskip-2cm
 \includegraphics[width=8cm, height=7cm]{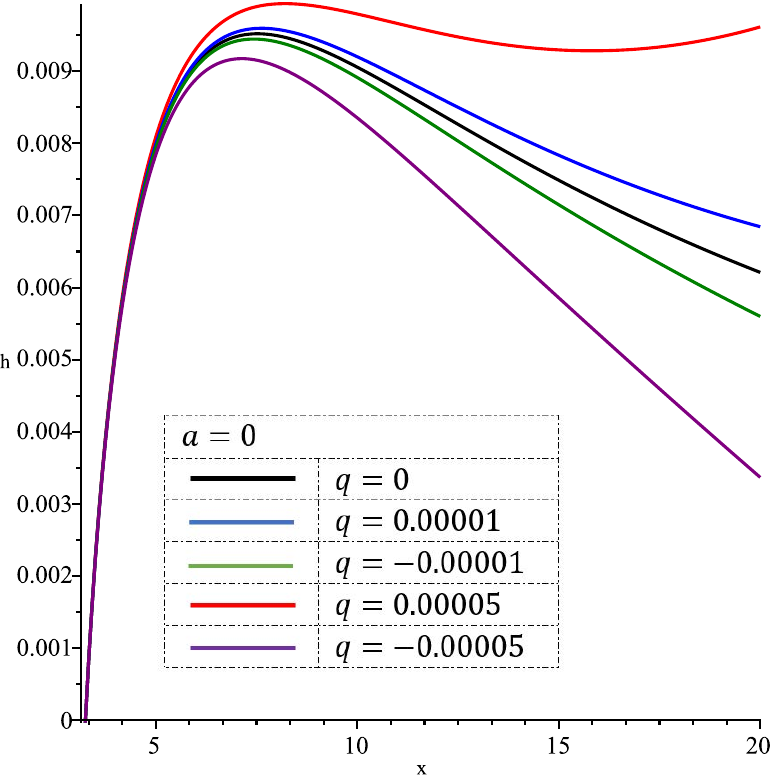}
 \captionof{figure}{Height scale h of the disk for \(q=0\) and \(q\neq0\) and \(a=0\). }
\end{Figure}

\begin{Figure}
 \centering
 \advance\leftskip-2cm
 \advance\rightskip-2cm
 \includegraphics[width=8cm, height=7cm]{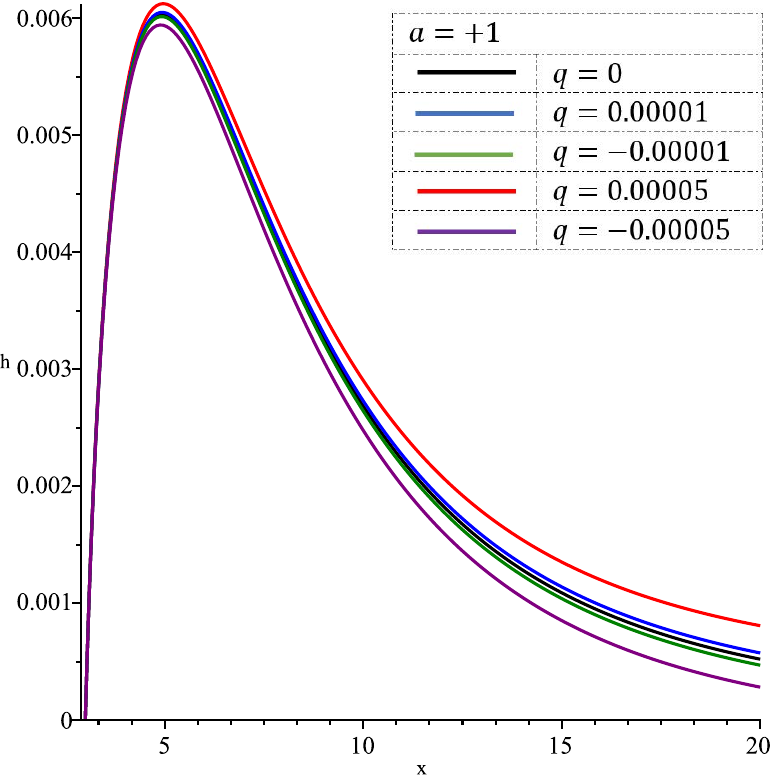}
 \captionof{figure}{Height scale h of the disk for \(q=0\) and \(q\neq0\) and \(a=+1\). }
\end{Figure}

Figures 5 and 6 illustrate the effect of the quadrupole moment on the shape of the black hole’s event horizon. The quadrupole moment distorts the event horizon from a perfect sphere, which in turn affects the shape and properties of the accretion disk. As the distance from the black hole increases, the influence of the quadrupole moment becomes more pronounced, leading to the observed deviations from the Kerr case. In the non-rotating scenario, we observe more significant changes, suggesting that rotation reduces the impact of the quadrupole moment, which appears more logical.

\begin{Figure}
 \centering
 \advance\leftskip-2cm
 \advance\rightskip-2cm
 \includegraphics[width=8cm, height=7cm]{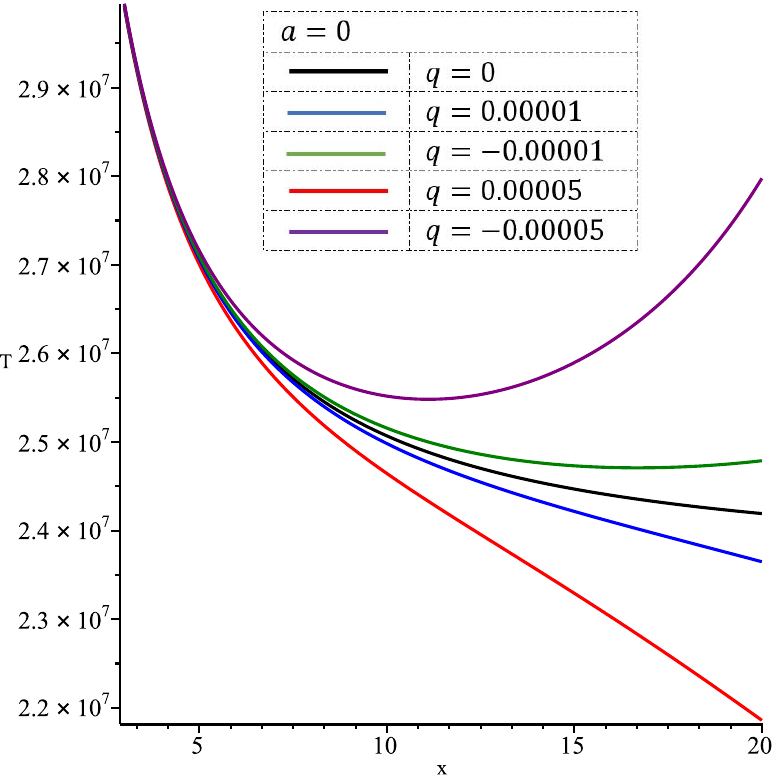}
 \captionof{figure}{Temperature for \(q=0\) and \(q\neq0\) in the unit k and \(a=0\).}
\end{Figure}

\begin{Figure}
 \centering
 \advance\leftskip-2cm
 \advance\rightskip-2cm
 \includegraphics[width=8cm, height=7cm]{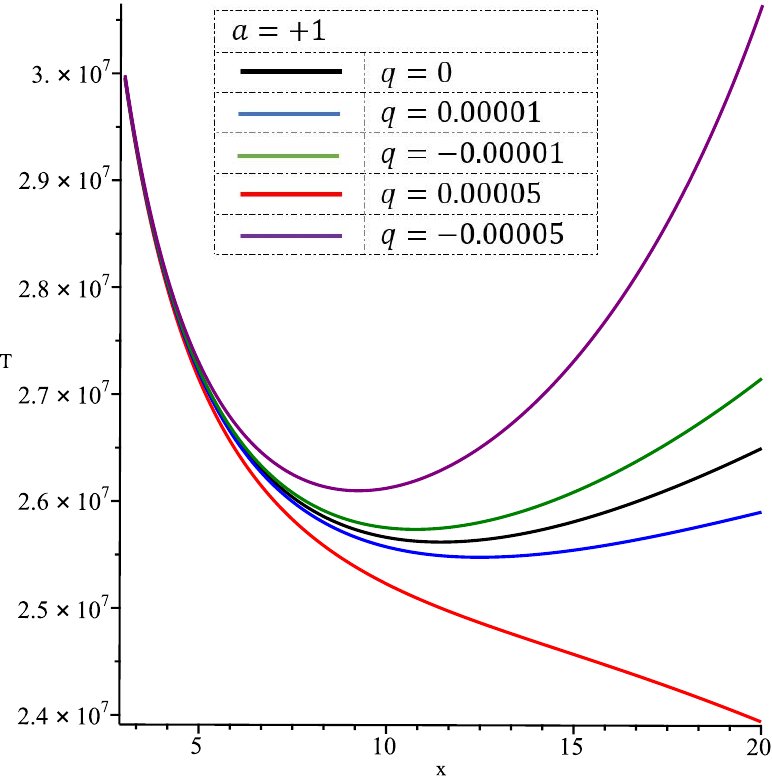}
 \captionof{figure}{Temperature for \(q=0\) and \(q\neq0\) in the unit k and \(a=+1\).}
\end{Figure}

In Figures 7 and 8, at small radii, the temperature difference across all cases is nearly the same and closely aligned. However, as the distance from the black hole increases, the quadrupole moment becomes more significant, affecting both the shape and orientation of the disk, which in turn influences the energy distribution. In the case of a positive quadrupole moment, there is a more rapid decrease in temperature at larger radii. Conversely, for a negative quadrupole moment, the outer regions receive more energy, resulting in a slower decrease in temperature at larger radii.
When rotation is introduced, the slope of the graph decreases.

\begin{Figure}
 \centering
 \advance\leftskip-2cm
 \advance\rightskip-2cm
 \includegraphics[width=8cm, height=7cm]{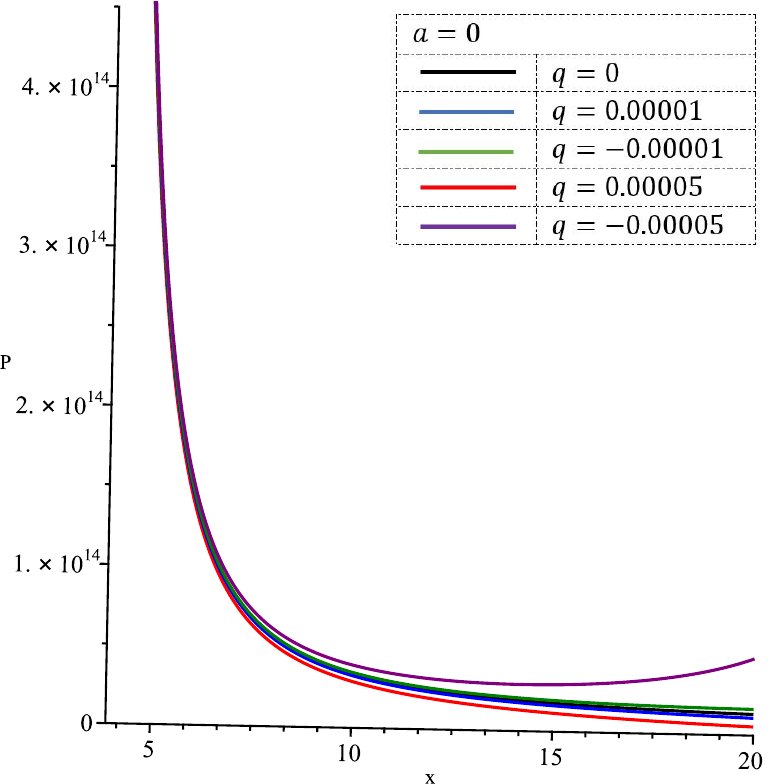}
 \captionof{figure}{Pressure for \(q=0\) and \(q\neq0\) in the $(\frac{dyn}{cm^2})$ unit for Non-rotating mode (a=0)}
\end{Figure}

\begin{Figure}
 \centering
 \advance\leftskip-2cm
 \advance\rightskip-2cm
 \includegraphics[width=8cm, height=7cm]{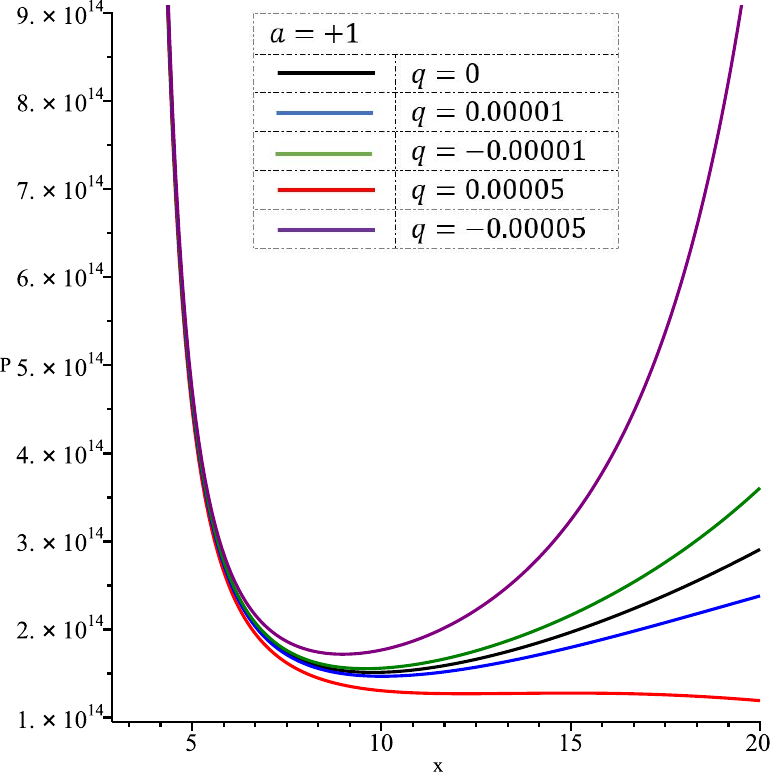}
 \captionof{figure}{Pressure for \(q=0\) and \(q\neq0\) in the $(\frac{dyn}{cm^2})$ unit and \(a=+1\).}
\end{Figure}

At a significant distance from the black hole, the pressure remains high and can be measured from approximately x = 5. Adding rotation makes pressure changes more noticeable. We expect the pressure graph in an accretion disk to start at a maximum, exhibit a certain slope down to a minimum, and show minimal fluctuations between these extreme. For this reason, cases where we observe multiple maxima cannot accurately explain the pressure graph.(See Figures 9 and 10)

The graphs that assess viscosity indicate that at the initial points, the viscosity is five times higher in the Kerr mode than in the Schwarzschild mode. Changes in the state of q = 0 are not noticeable; however, in states with and without rotation, as well as with the addition of a quadrupole, the slopes of the graphs change. (See Figure 11 and 12)

\begin{Figure}
 \centering
 \advance\leftskip-2cm
 \advance\rightskip-2cm
 \includegraphics[width=8cm, height=7cm]{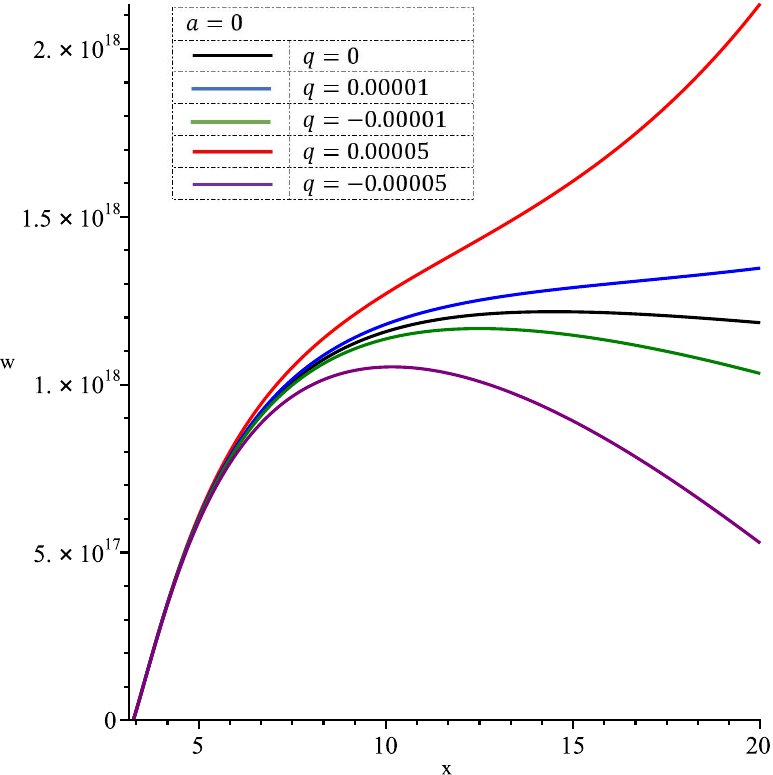}
 \captionof{figure}{viscosity for \(q=0\) and \(q\neq0\) in the $(\frac{dyn}{cm})$ unit and \(a=0\)}
\end{Figure}

\begin{Figure}
 \centering
 \advance\leftskip-2cm
 \advance\rightskip-2cm
 \includegraphics[width=8cm, height=7cm]{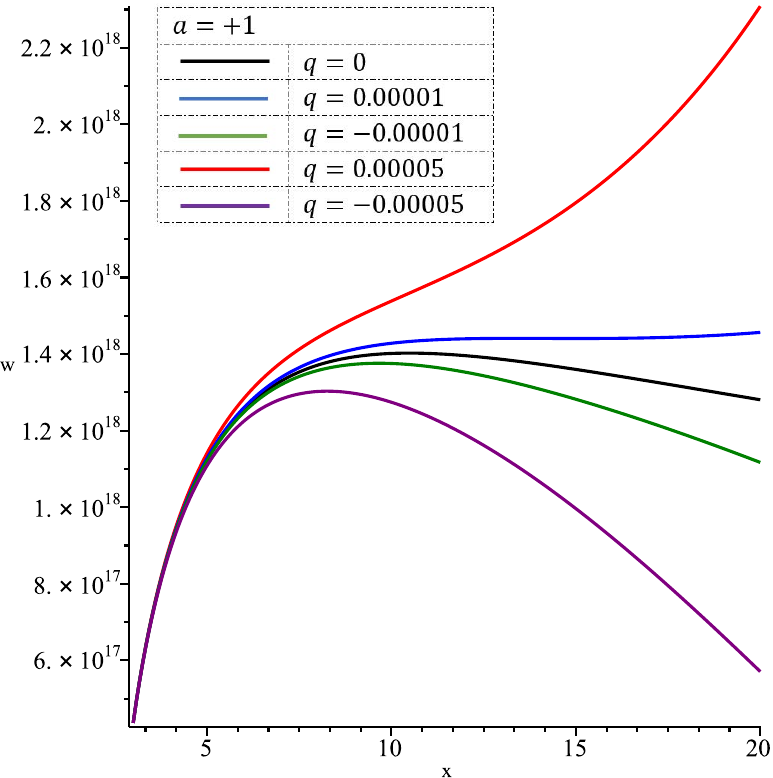}
 \captionof{figure}{viscosity for \(q=0\) and \(q\neq0\) in the $(\frac{dyn}{cm})$ unit and \(a=+1\).}
\end{Figure}

Considering a fixed value of q, we can examine how changes in rotation with a quadrupole affect each figure. As we gradually increase the rotation, the variations in the figures will indicate how the desired quantity changes in different situations.

\begin{Figure}
 \centering
 \advance\leftskip-2cm
 \advance\rightskip-2cm
 \includegraphics[width=8cm, height=7cm]{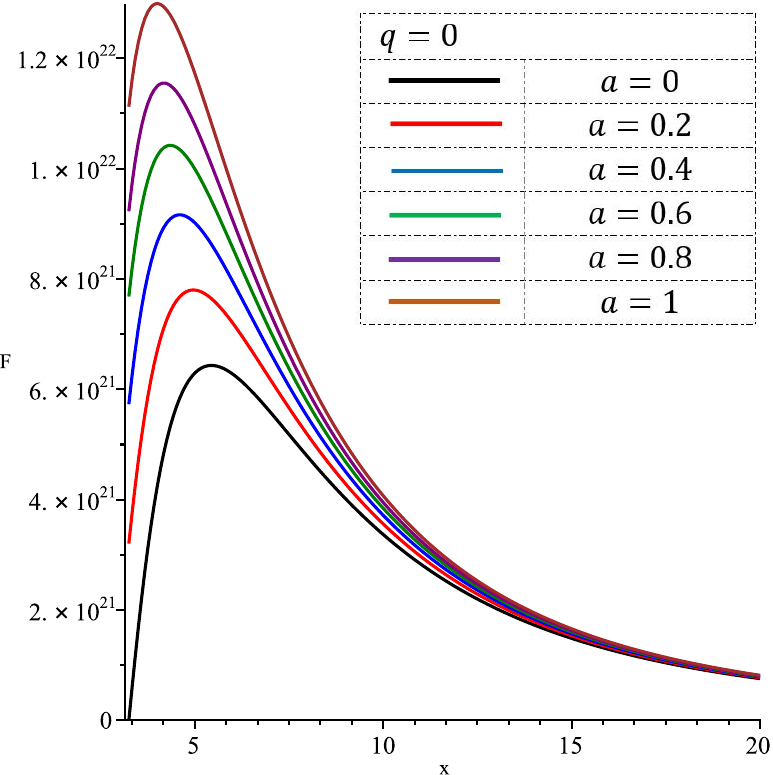}
 \captionof{figure}{Flux for \(a=0\) to \(a=1\) where \(q=0\).}
\end{Figure}

\begin{Figure}
 \centering
 \advance\leftskip-2cm
 \advance\rightskip-2cm
 \includegraphics[width=8cm, height=7cm]{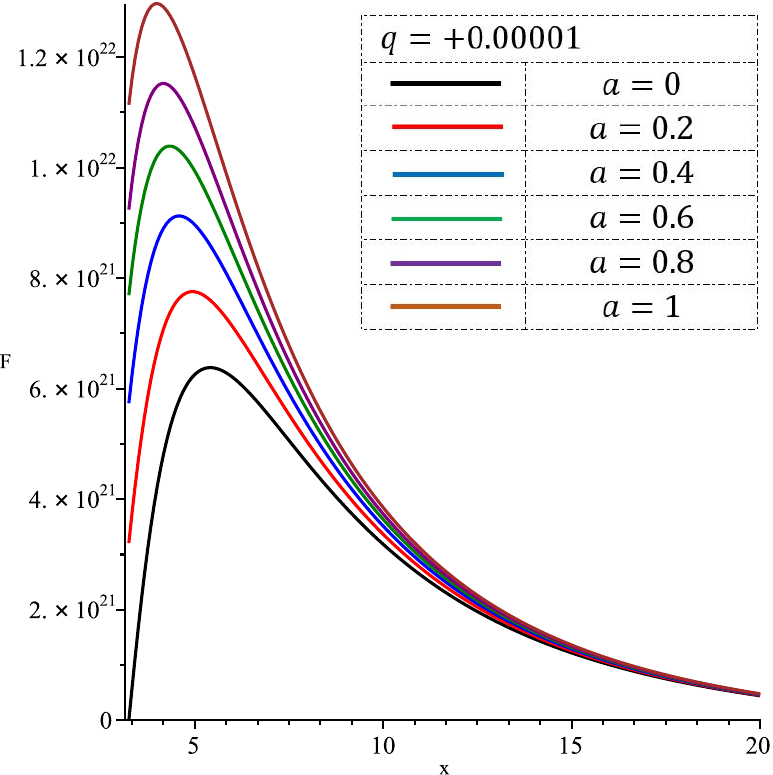}
 \captionof{figure}{Flux for \(a=0\) to \(a=1\) where \(q=0.00001\).}
\end{Figure}

\begin{Figure}
 \centering
 \advance\leftskip-2cm
 \advance\rightskip-2cm
 \includegraphics[width=8cm, height=7cm]{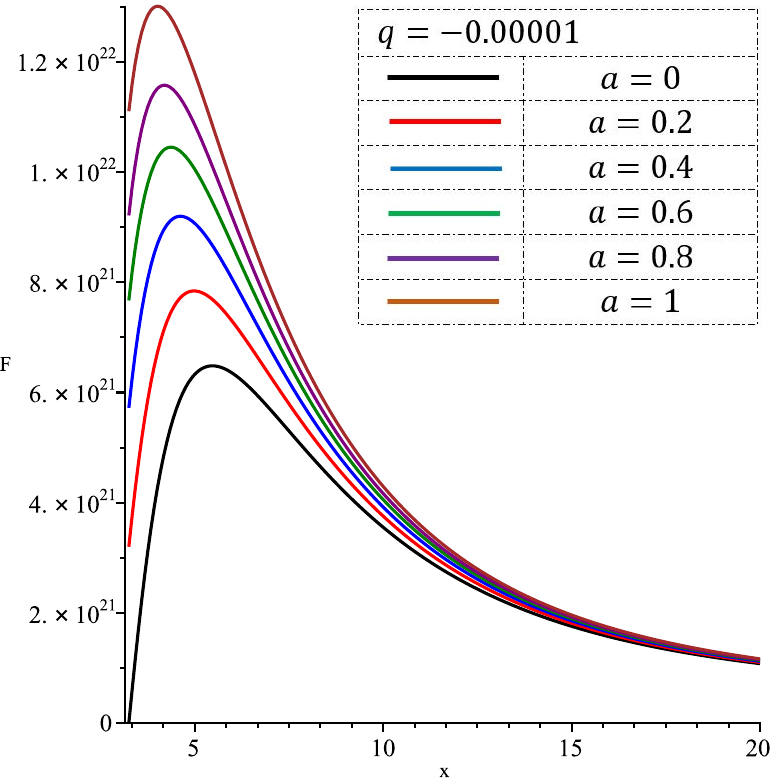}
 \captionof{figure}{Flux for \(a=0\) to \(a=1\) where \(q=-0.00001\).}
\end{Figure}

In Figures 13, 14, and 15, we explore three values of q accepted for increasingly detailed investigation. As we gradually add rotation to the disk, the peak of the figure (maximum flux) begins to rise, particularly as x (the location of the increase) decreases. This observation highlights the significance of rotation.

\begin{Figure}
 \centering
 \advance\leftskip-2cm
 \advance\rightskip-2cm
 \includegraphics[width=8cm, height=7cm]{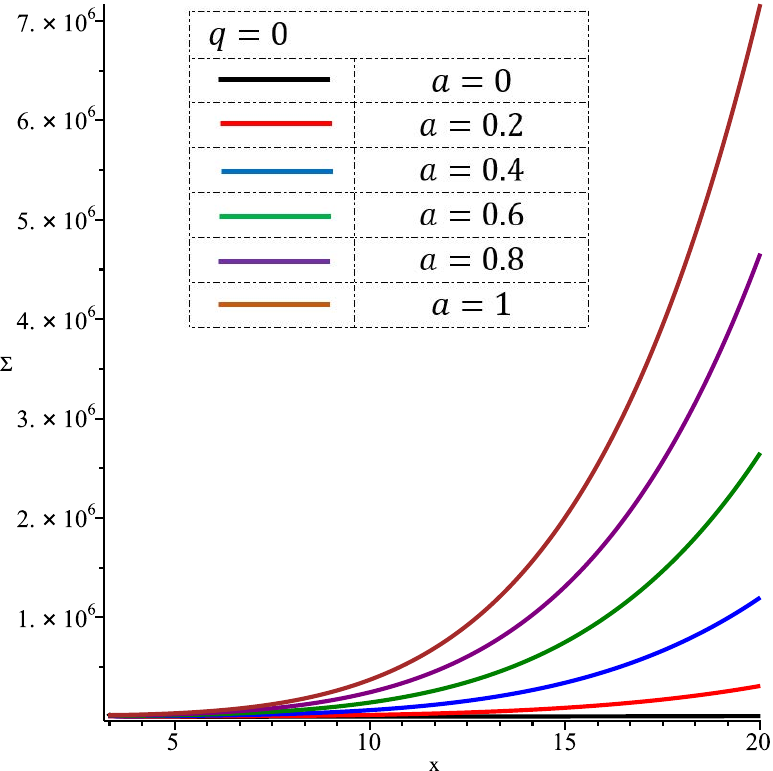}
 \captionof{figure}{surface density for \(a=0\) to \(a=1\) where \(q=0\).}
\end{Figure}

\begin{Figure}
 \centering
 \advance\leftskip-2cm
 \advance\rightskip-2cm
 \includegraphics[width=8cm, height=7cm]{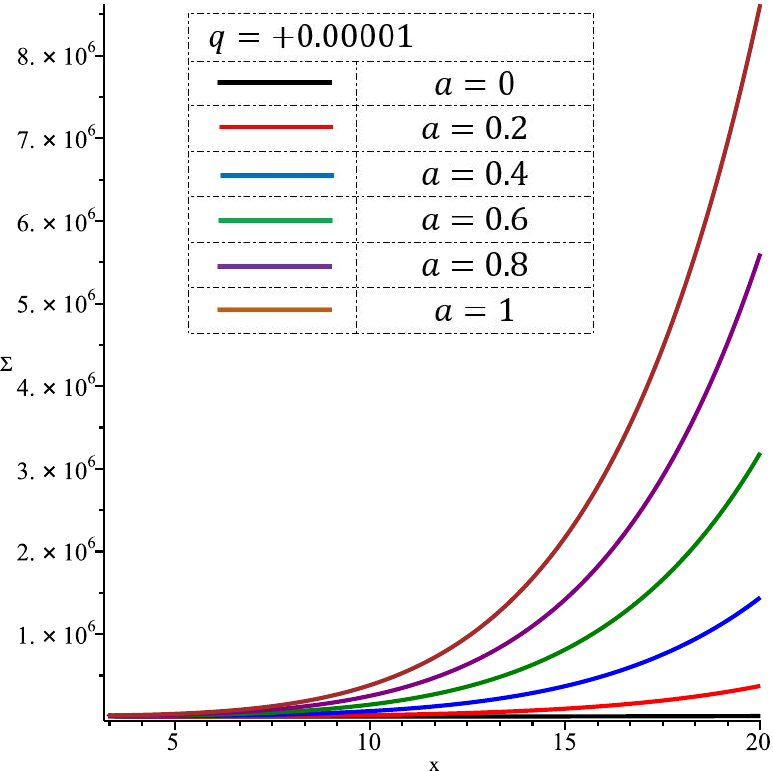}
 \captionof{figure}{surface density for \(a=0\) to \(a=1\) where \(q=0.00001\).}
\end{Figure}

\begin{Figure}
 \centering
 \advance\leftskip-2cm
 \advance\rightskip-2cm
 \includegraphics[width=8cm, height=7cm]{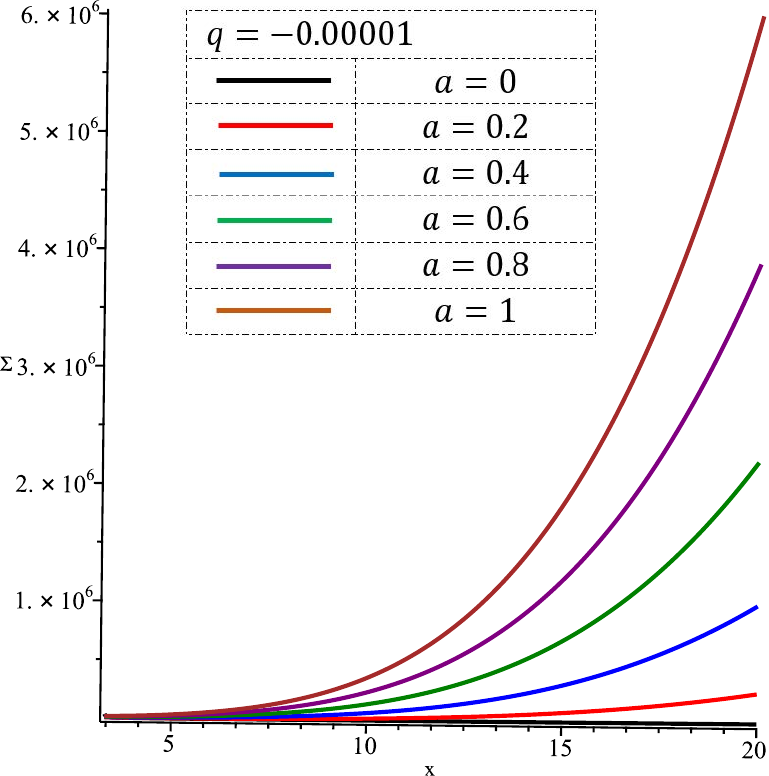}
 \captionof{figure}{surface density for \(a=0\) to \(a=1\) where \(q=-0.00001\).}
\end{Figure}

When discussing surface density, we observe that at a = 0, changes are minimal in for every three values of q (See Figures 16, 17, and 18). However, with increasing rotation, the value of x rises rapidly with distance from the black hole. This observation warrants further investigations to correlate this state with real samples. Moreover, there is a stronger agreement that an analytical model can be aligned with a real model to yield better results.\\
The figures indicate that for positive q, the surface density increases at a steeper slope, while for negative q, the slope is shallower in two similar cases compared to q = 0.

\begin{Figure}
 \centering
 \advance\leftskip-2cm
 \advance\rightskip-2cm
 \includegraphics[width=8cm, height=7cm]{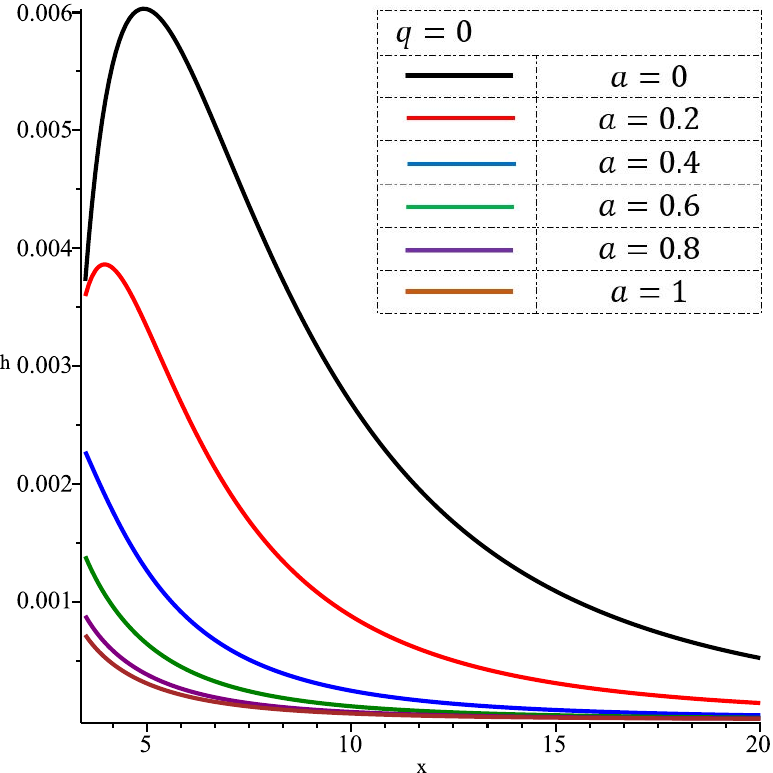}
 \captionof{figure}{Height scale for \(a=0\) to \(a=1\) where \(q=0\).}
\end{Figure}

\begin{Figure}
 \centering
 \advance\leftskip-2cm
 \advance\rightskip-2cm
 \includegraphics[width=8cm, height=7cm]{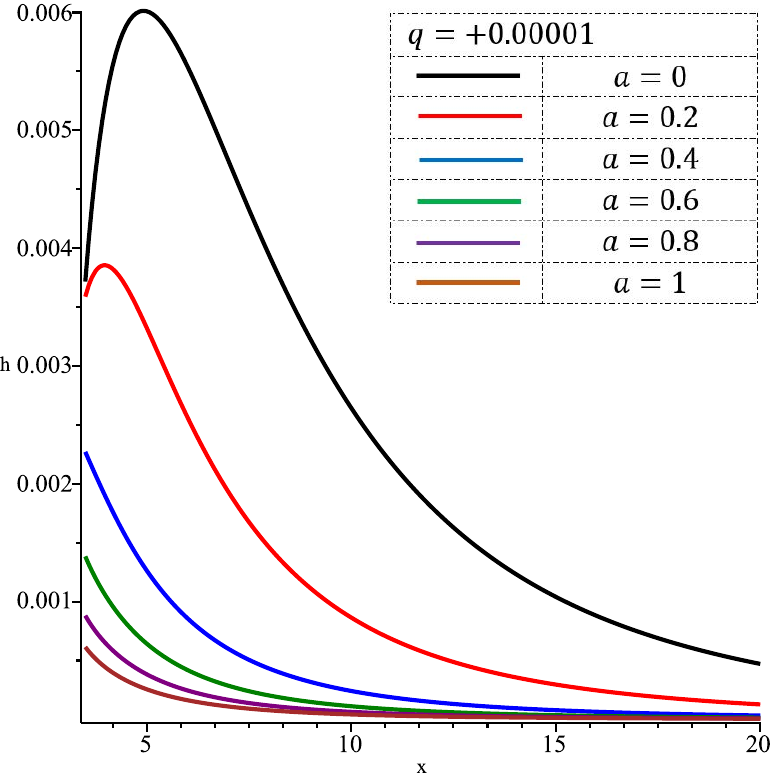}
 \captionof{figure}{Height scale for \(a=0\) to \(a=1\) where \(q=0.00001\).}
\end{Figure}

\begin{Figure}
 \centering
 \advance\leftskip-2cm
 \advance\rightskip-2cm
 \includegraphics[width=8cm, height=7cm]{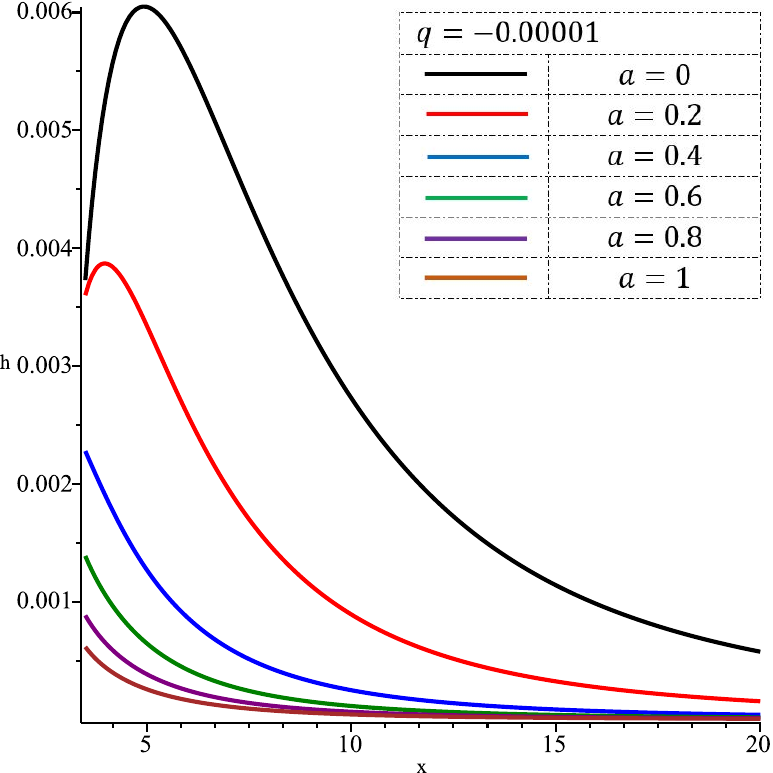}
 \captionof{figure}{Height scale for \(a=0\) to \(a=1\) where \(q=-0.00001\).}
\end{Figure}

When comparing height scales in the non-rotating Schwarzschild state to those in the rotating Kerr state, a significant difference emerges. As the rotation parameter (a) increases, the figures representing the height scales approach the horizon. Consequently, the differences between the various values of quadrpole moments (q) become negligible, and ultimately, the figures overlap effectively  as the distance from the horizon (x) increases. (See Figures 19, 20, and 21)

\begin{Figure}
 \centering
 \advance\leftskip-2cm
 \advance\rightskip-2cm
 \includegraphics[width=8cm, height=7cm]{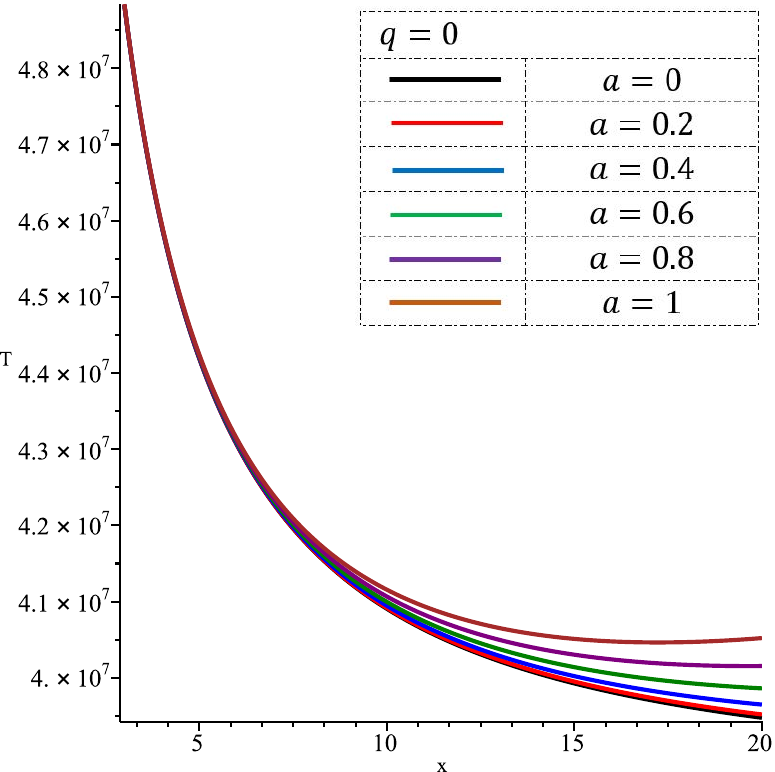}
 \captionof{figure}{Temperature for \(a=0\) to \(a=1\) where \(q=0\).}
\end{Figure}

\begin{Figure}
 \centering
 \advance\leftskip-2cm
 \advance\rightskip-2cm
 \includegraphics[width=8cm, height=7cm]{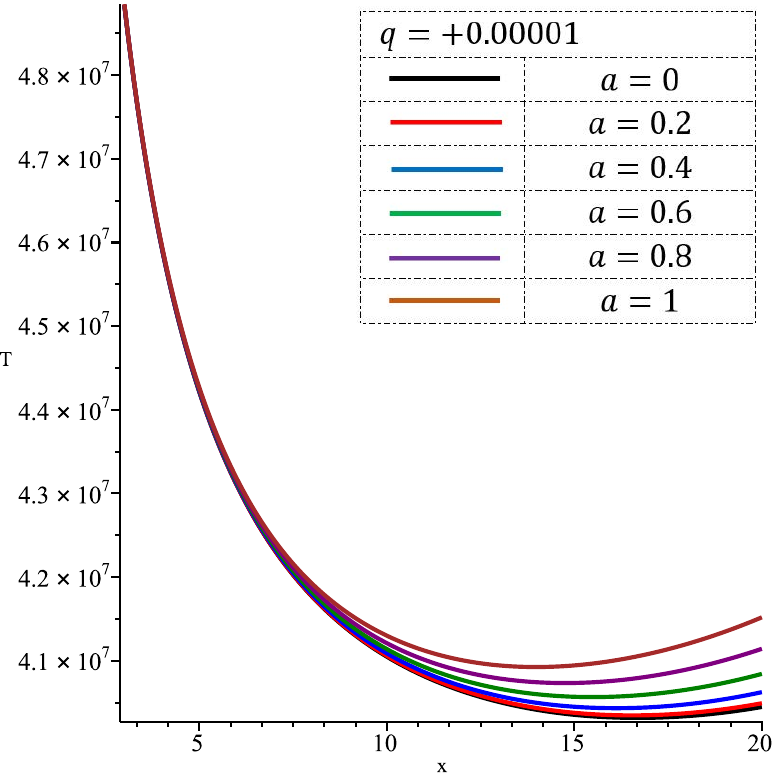}
 \captionof{figure}{Temperature for \(a=0\) to \(a=1\) where \(q=0.00001\).}
\end{Figure}

\begin{Figure}
 \centering
 \advance\leftskip-2cm
 \advance\rightskip-2cm
 \includegraphics[width=8cm, height=7cm]{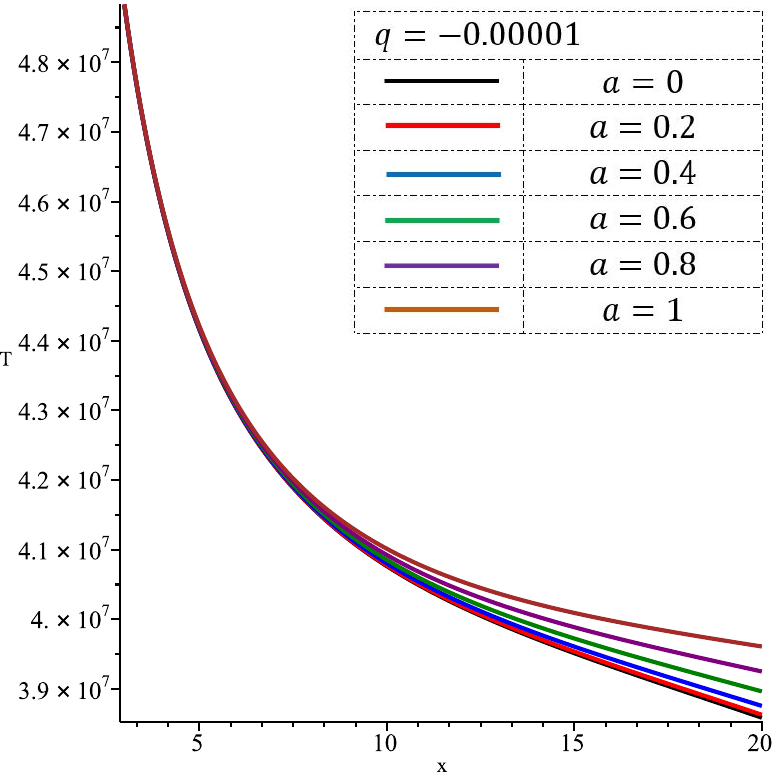}
 \captionof{figure}{Temperature for \(a=0\) to \(a=1\) where \(q=-0.00001\).}
\end{Figure}

Examining the temperature graphs in Figures 22, 23, and 24 reveals that changes in the rotation parameter (a) significantly affect temperature across  different values of q. Specifically, in q = 0.00001, increasing rotation leads to a substantial change in temperature, while in q = -0.00001, the temperature difference between no rotation and low rotation is negligible.

\begin{Figure}
 \centering
 \advance\leftskip-2cm
 \advance\rightskip-2cm
 \includegraphics[width=8cm, height=7cm]{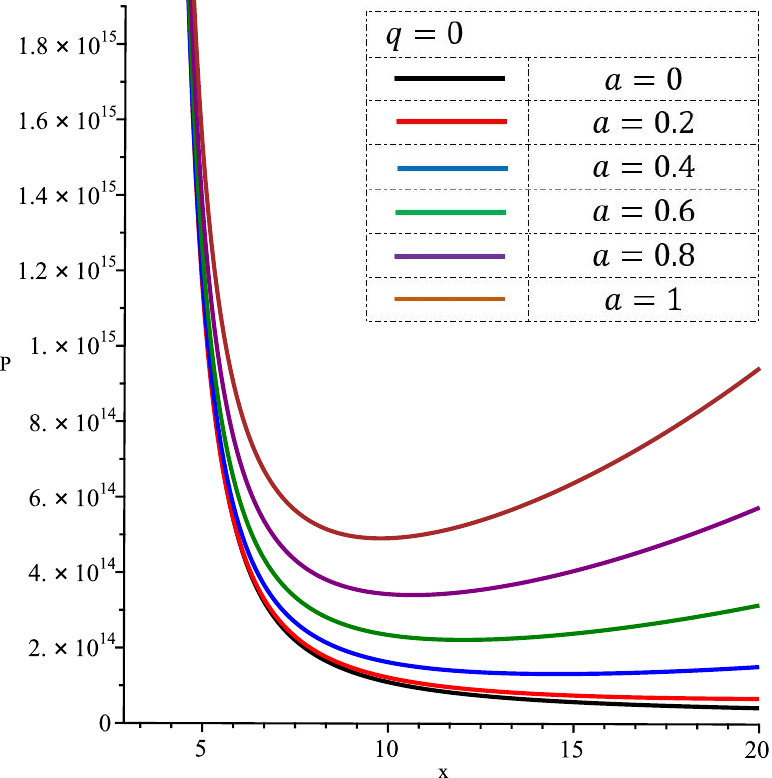}
 \captionof{figure}{Pressure for \(a=0\) to \(a=1\) where \(q=0\).}
\end{Figure}

\begin{Figure}
 \centering
 \advance\leftskip-2cm
 \advance\rightskip-2cm
 \includegraphics[width=8cm, height=7cm]{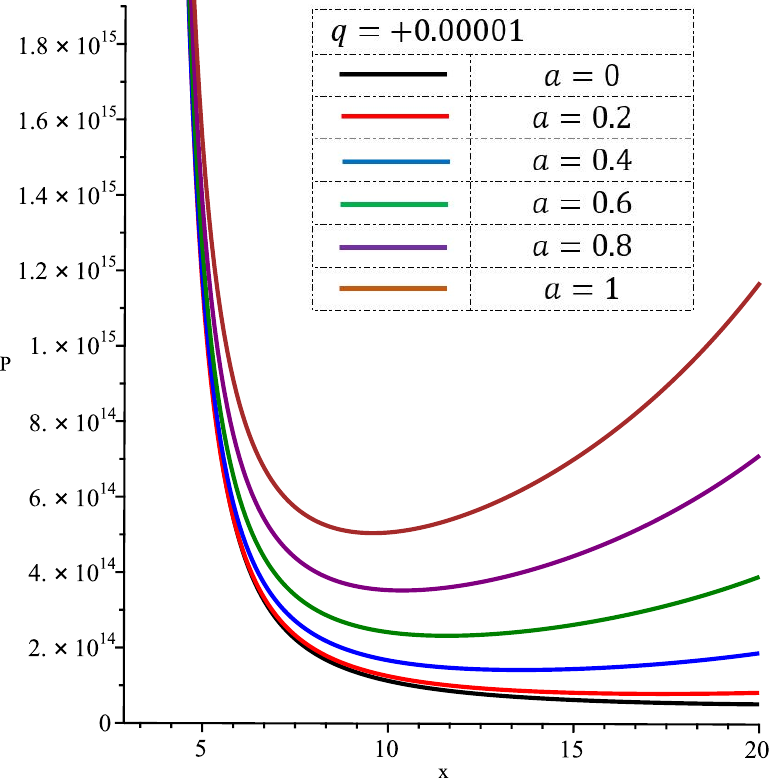}
 \captionof{figure}{Pressure for \(a=0\) to \(a=1\) where \(q=0.00001\).}
\end{Figure}

\begin{Figure}
 \centering
 \advance\leftskip-2cm
 \advance\rightskip-2cm
 \includegraphics[width=8cm, height=7cm]{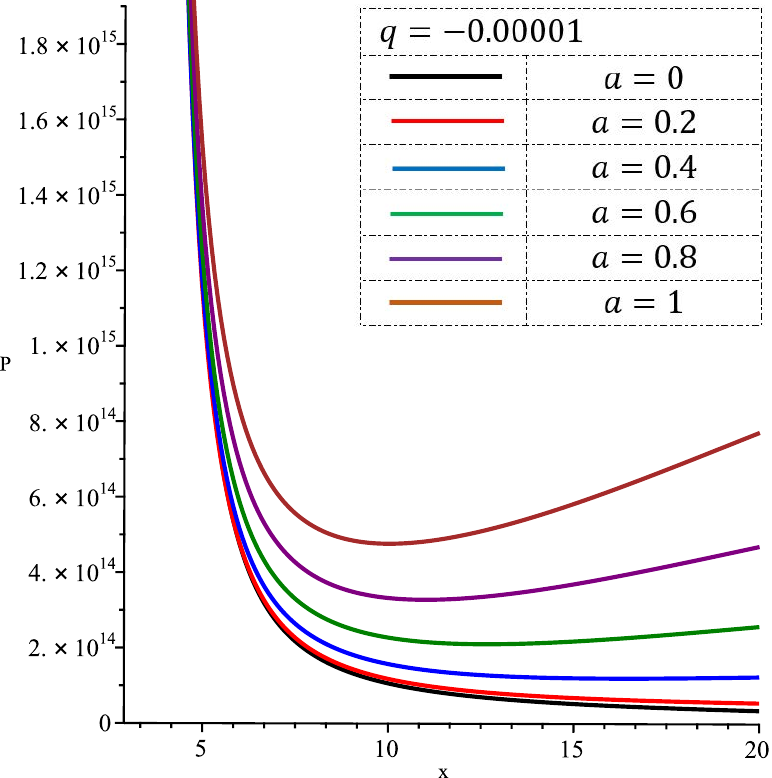}
 \captionof{figure}{Pressure for \(a=0\) to \(a=1\) where \(q=-0.00001\).}
\end{Figure}

When examining pressure changes for values of (a) ranging from 0 to 1, we observe a general trend of decreasing pressure as we move away from the origin. However, in some instances, we see an unexpected increase in pressure, which deviates from what would be expected in real-world scenarios. (See Figures 25, 26, and 27). 

\begin{Figure}
 \centering
 \advance\leftskip-2cm
 \advance\rightskip-2cm
 \includegraphics[width=8cm, height=7cm]{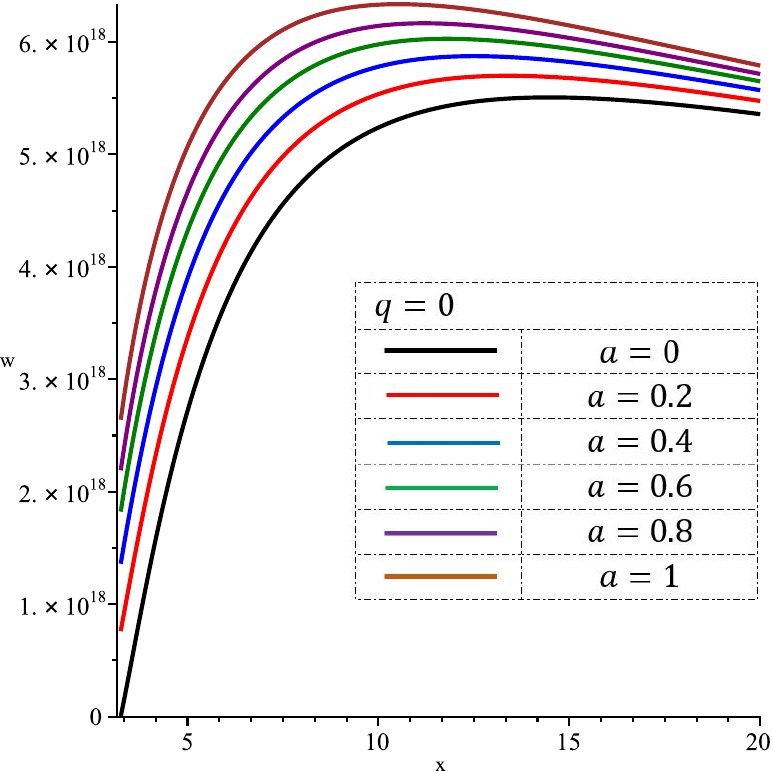}
 \captionof{figure}{viscous stress for \(a=0\) to \(a=1\) where \(q=0\).}
\end{Figure}

\begin{Figure}
 \centering
 \advance\leftskip-2cm
 \advance\rightskip-2cm
 \includegraphics[width=8cm, height=7cm]{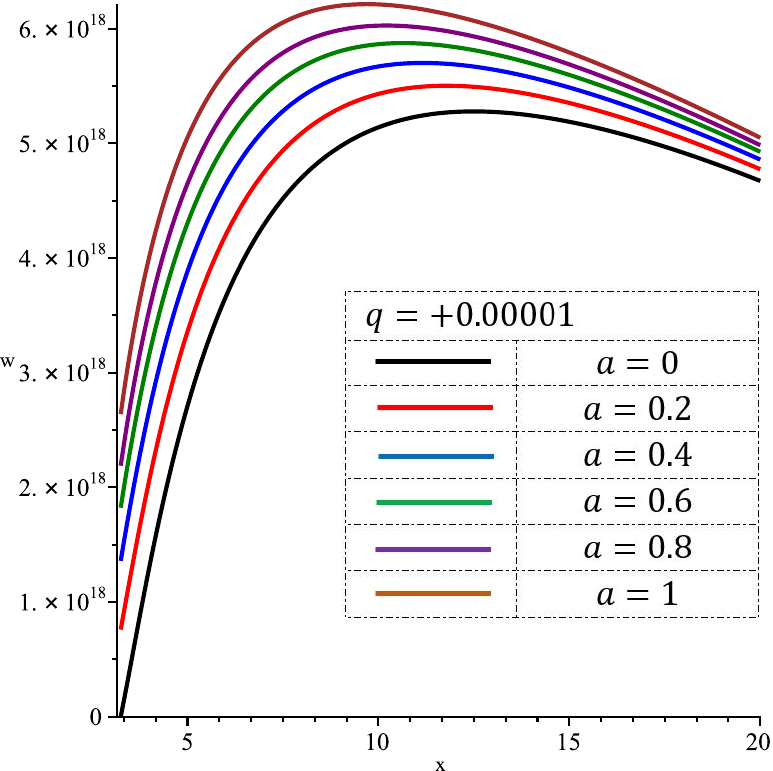}
 \captionof{figure}{viscous stress for \(a=0\) to \(a=1\) where \(q=0.00001\).}
\end{Figure}

\begin{Figure}
 \centering
 \advance\leftskip-2cm
 \advance\rightskip-2cm
 \includegraphics[width=8cm, height=7cm]{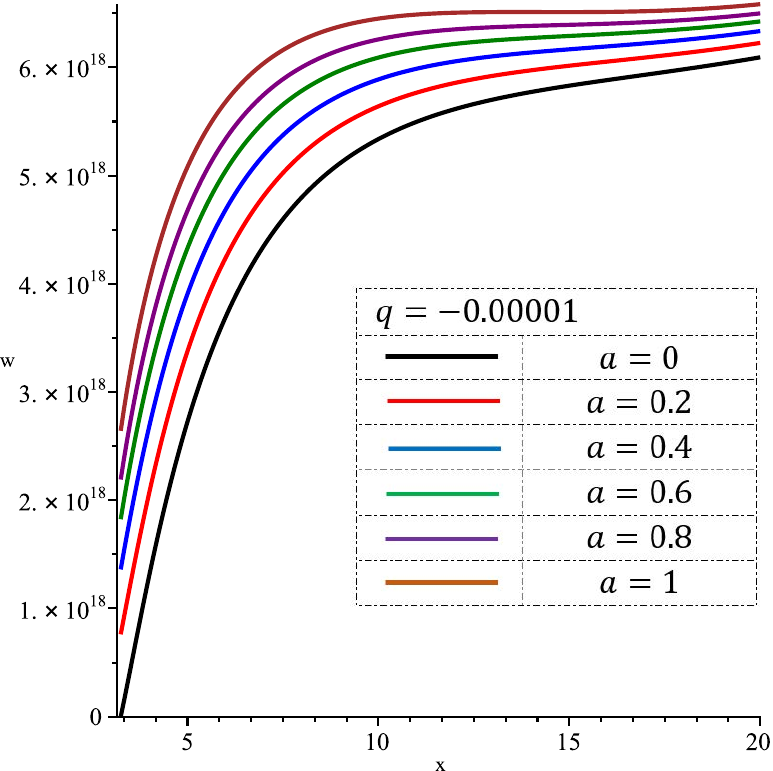}
 \captionof{figure}{viscous stress for \(a=0\) to \(a=1\) where \(q=-0.00001\).}
\end{Figure}

Figures 28 to 30 illustrate the effects of rotation on viscosity for various values of q. In case of q = 0, increasing rotation exhibits a predictable behavior. However, at negative values of q, the observed behavior deviates significantly from expectations. In contrast, increasing the rotation parameter (a) at positive values of q yields a more realistic representation of the changing viscosity.\\

Our analysis focuses on understanding the impact of varying the parameter \(\alpha\)\ on the results within a quadruple system. By systematically adjusting \(\alpha\)\ across different modes and intervals, we can observe the resulting variations and gain insights into the system's behavior. Previously, \(\alpha\)\ was held constant at a value of 0.1. Now, by varying \(\alpha\)\ while keeping q and a constant, we obtain intriguing graphs that reveal valuable information. This approach allows us to comprehensively assess the influence of \(\alpha\)\ on the results and draw meaningful conclusions relevant to future studies. For this investigation, \(\alpha\)\ is varied across three modes, with intervals of 0.05.

\begin{Figure}
 \centering
 \advance\leftskip-2cm
 \advance\rightskip-2cm
 \includegraphics[width=8cm, height=6.9cm]{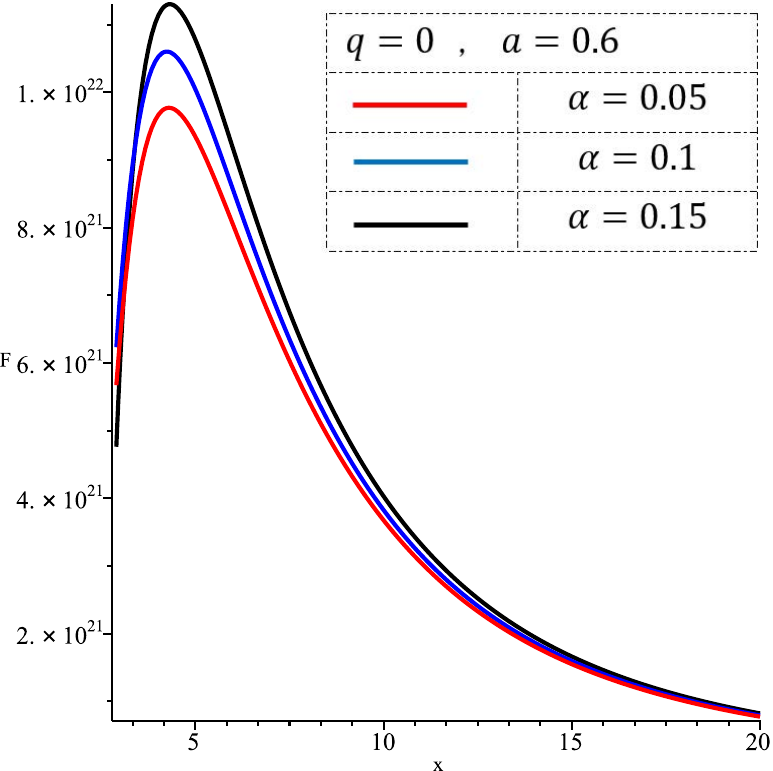}
 \captionof{figure}{Flux for \(\alpha=0.01\) , \(\alpha=0.1\) and \(\alpha=0.5\) where \(q=0\).}
\end{Figure}

\begin{Figure}
 \centering
 \advance\leftskip-2cm
 \advance\rightskip-2cm
 \includegraphics[width=8cm, height=6.9cm]{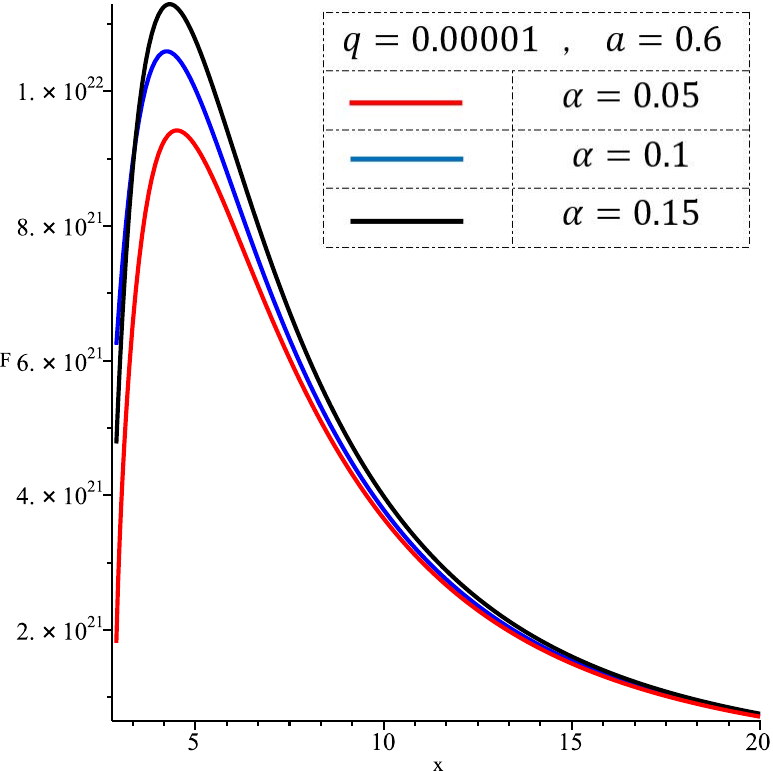}
 \captionof{figure}{Flux for \(\alpha=0.01\) , \(\alpha=0.1\) and \(\alpha=0.5\) where \(q=0.00001\).}
\end{Figure}

\begin{Figure}
 \centering
 \advance\leftskip-2cm
 \advance\rightskip-2cm
 \includegraphics[width=8cm, height=6.9cm]{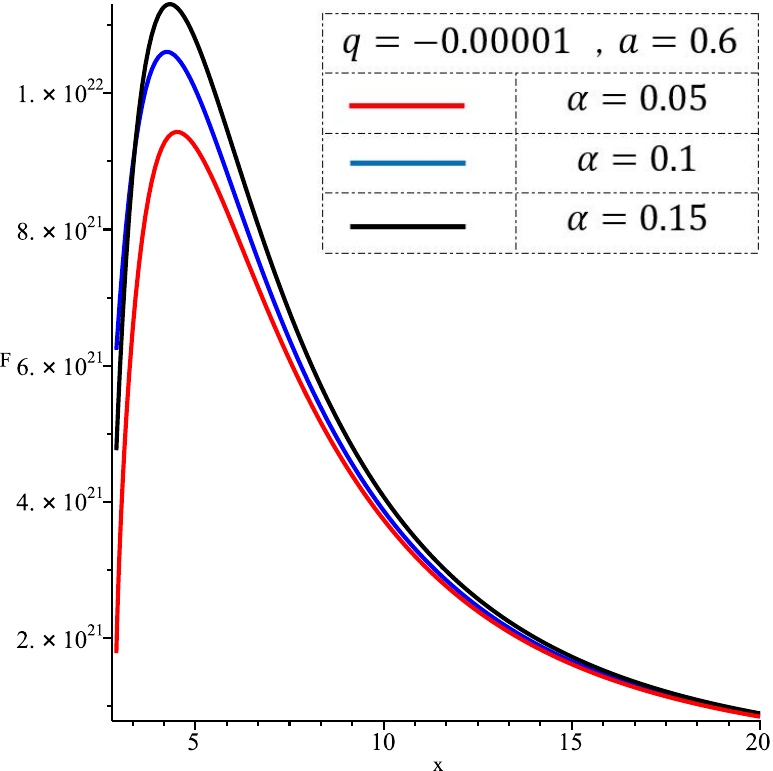}
 \captionof{figure}{Flux for \(\alpha=0.01\) , \(\alpha=0.1\) and \(\alpha=0.5\) where \(q=-0.00001\).}
\end{Figure}

Figures 31, 32, and 33 demonstrate the effects of both \(\alpha\)\ and quadrupole on flux. We observe a consistent pattern: as \(\alpha\)\ increases, the flux experiences a noticeable rise at the initial points, culminating in pronounced peaks. While \(\alpha\)\ significantly influences the initial flux behavior and peak values, the overall flux level converges to a similar shape and value across different \(\alpha\)\ values, suggesting stabilization at larger distances from the horizon. \\
Furthermore, our analysis reveals a significant impact of the quadrupole (both positive and negative) on the initial flux, characterized by a steeper slope compared to q = 0. This indicates that the quadrupole effect accelerates flux growth near the horizon points.

\begin{Figure}
 \centering
 \advance\leftskip-2cm
 \advance\rightskip-2cm
 \includegraphics[width=8cm, height=7cm]{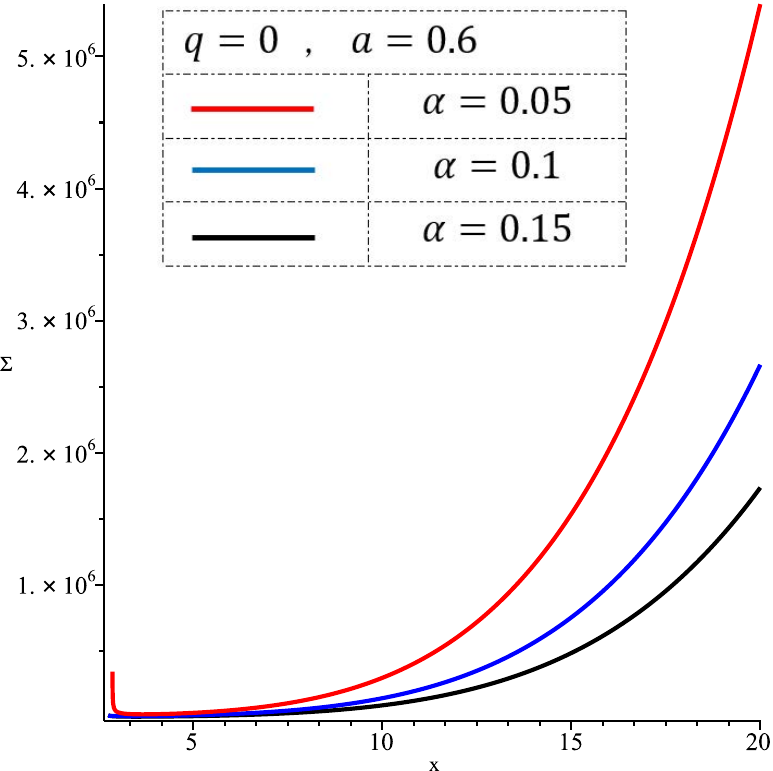}
 \captionof{figure}{surface density for \(\alpha=0.01\) , \(\alpha=0.1\) and \(\alpha=0.5\) where \(q=0\).}
\end{Figure}

\begin{Figure}
 \centering
 \advance\leftskip-2cm
 \advance\rightskip-2cm
 \includegraphics[width=8cm, height=7cm]{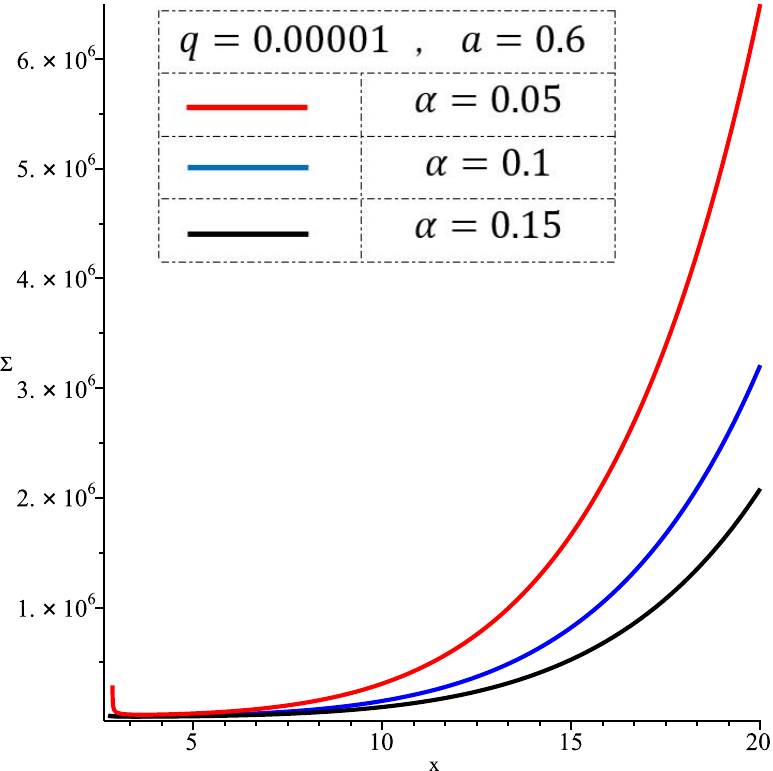}
 \captionof{figure}{surface density for \(\alpha=0.01\) , \(\alpha=0.1\) and \(\alpha=0.5\) where \(q=0.00001\).}
\end{Figure}

\begin{Figure}
 \centering
 \advance\leftskip-2cm
 \advance\rightskip-2cm
 \includegraphics[width=8cm, height=7cm]{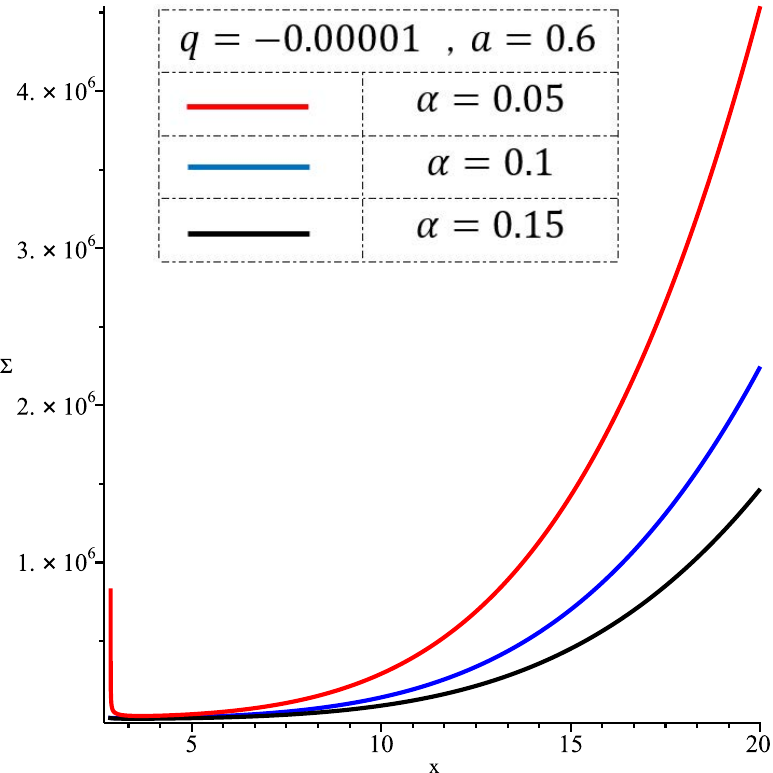}
 \captionof{figure}{surface density for \(\alpha=0.01\) , \(\alpha=0.1\) and \(\alpha=0.5\) where \(q=-0.00001\).}
\end{Figure}

Applying a positive quadrupole results in a steeper slope of the surface density profile at certain distances from the black hole, while a negative quadrupole leads to a more gradual decrease in slope (Figures 34, 35, and 36). These observations highlight the significant influence of the quadrupole on the surface density distribution as we move away from the black hole. By analyzing these slope changes, we can gain insights into how the quadrupole affects the distribution of mass or matter density at different distances. Further investigation and analysis of these patterns would be valuable to understand the underlying physical mechanisms driving these density variations with respect to the quadrupole. Delving deeper into these relationships could provide crucial insights into the behavior of the system under study.

\begin{Figure}
 \centering
 \advance\leftskip-2cm
 \advance\rightskip-2cm
 \includegraphics[width=8cm, height=6.9cm]{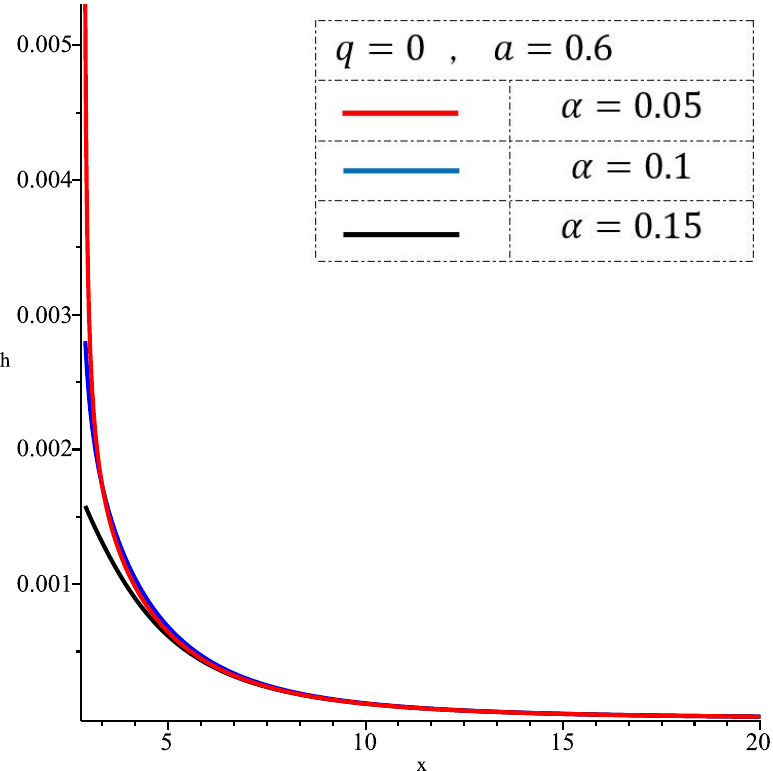}
 \captionof{figure}{Height scale for \(\alpha=0.01\) , \(\alpha=0.1\) and \(\alpha=0.5\) where \(q=0\).}
\end{Figure}

\begin{Figure}
 \centering
 \advance\leftskip-2cm
 \advance\rightskip-2cm
 \includegraphics[width=8cm, height=6.9cm]{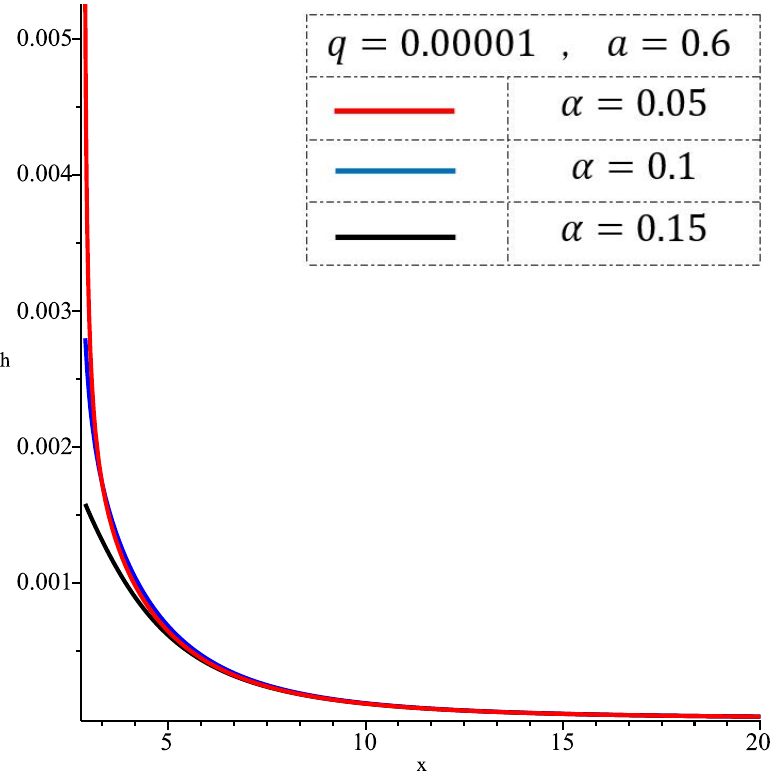}
 \captionof{figure}{Height scale for \(\alpha=0.01\) , \(\alpha=0.1\) and \(\alpha=0.5\) where \(q=0.00001\).}
\end{Figure}

\begin{Figure}
 \centering
 \advance\leftskip-2cm
 \advance\rightskip-2cm
 \includegraphics[width=8cm, height=6.9cm]{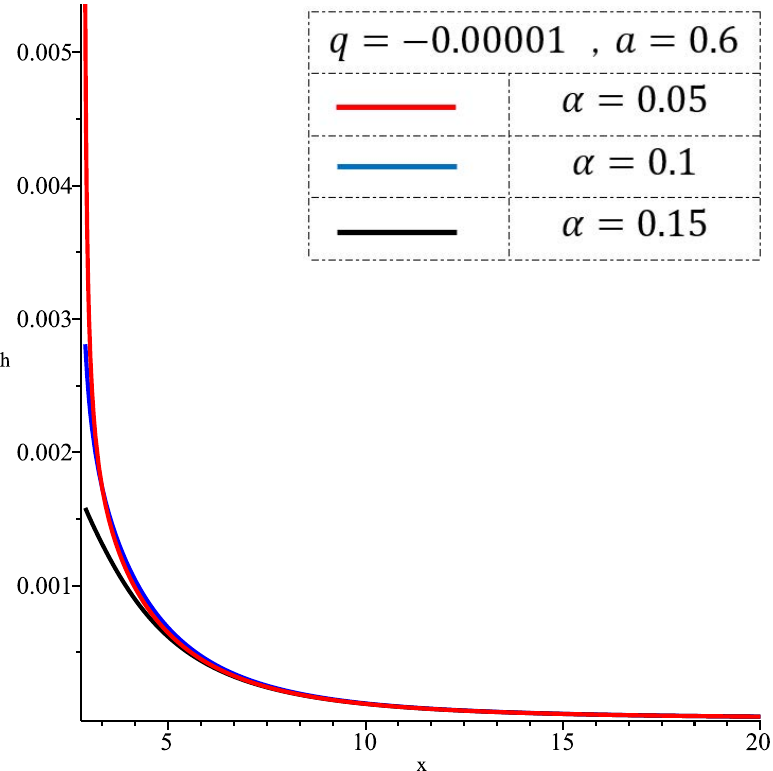}
 \captionof{figure}{Height scale for \(\alpha=0.01\) , \(\alpha=0.1\) and \(\alpha=0.5\) where \(q=-0.00001\).}
\end{Figure}

Our analysis of Figures 37, 38, and 39 indicates that variations in \(\alpha\)\ and quadrupole moment (q) have minimal impact on the h-x profiles. This suggests that the system is currently operating at a relatively stable state with respect to these parameters, as changes in \(\alpha\)\ have minimal effect on the overall behavior.

\begin{Figure}
 \centering
 \advance\leftskip-2cm
 \advance\rightskip-2cm
 \includegraphics[width=8cm, height=6.9cm]{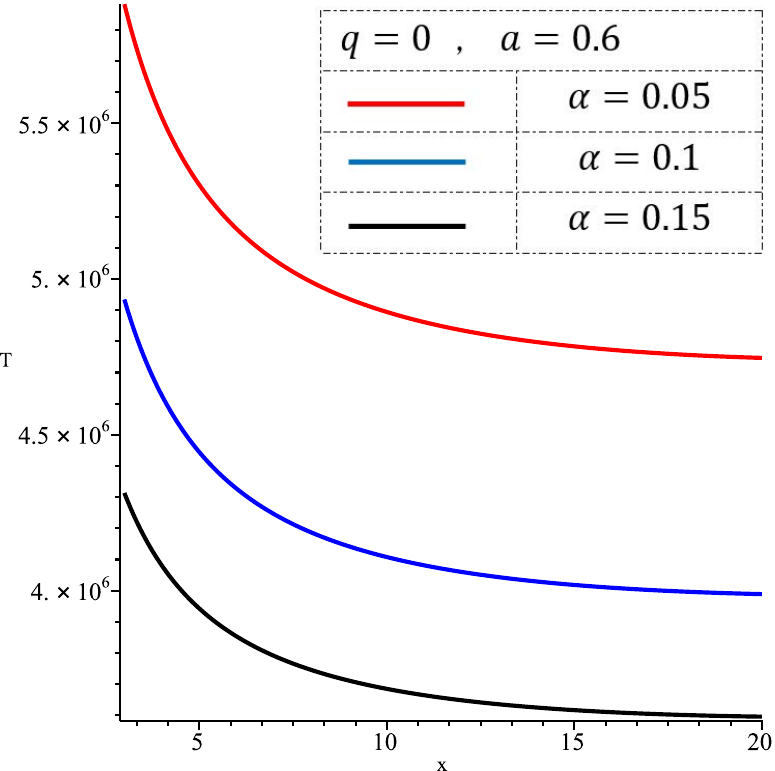}
 \captionof{figure}{Temperature for \(\alpha=0.01\) , \(\alpha=0.1\) and \(\alpha=0.5\) where \(q=0\).}
\end{Figure}

\begin{Figure}
 \centering
 \advance\leftskip-2cm
 \advance\rightskip-2cm
 \includegraphics[width=8cm, height=6.9cm]{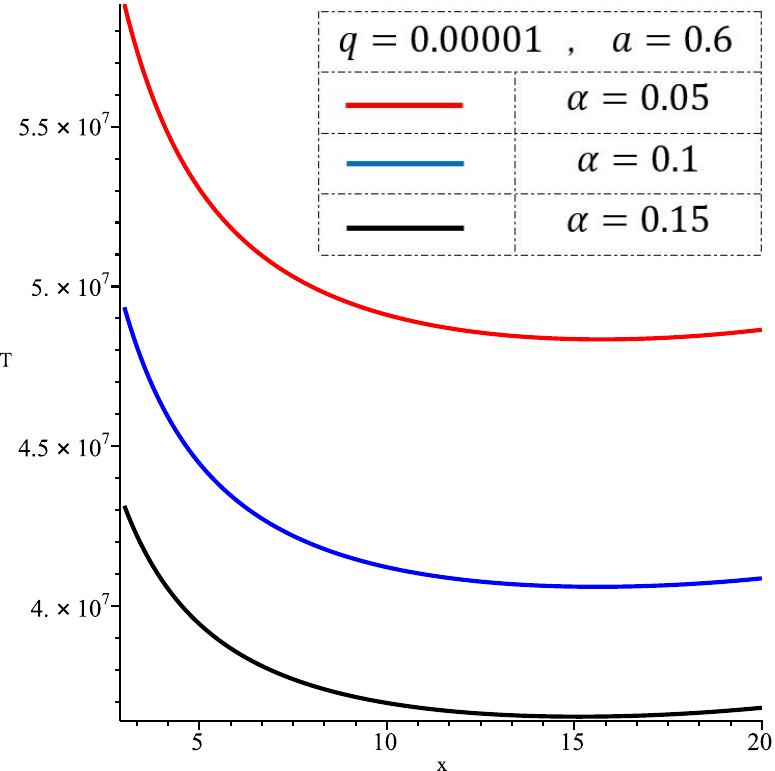}
 \captionof{figure}{Temperature for \(\alpha=0.01\) , \(\alpha=0.1\) and \(\alpha=0.5\) where \(q=0.00001\).}
\end{Figure}

\begin{Figure}
 \centering
 \advance\leftskip-2cm
 \advance\rightskip-2cm
 \includegraphics[width=8cm, height=6.9cm]{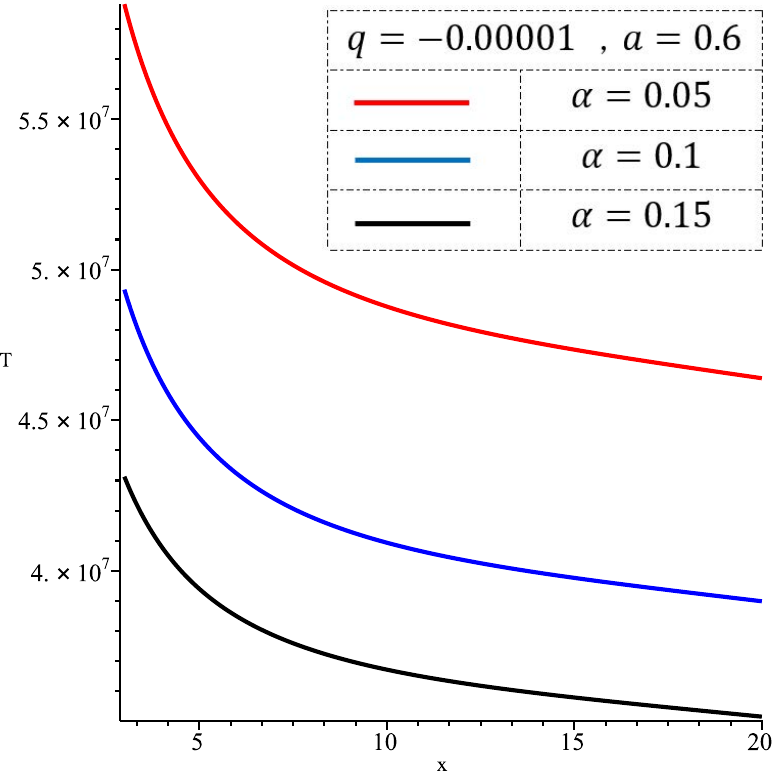}
 \captionof{figure}{Temperature for \(\alpha=0.01\) , \(\alpha=0.1\) and \(\alpha=0.5\) where \(q=-0.00001\).}
\end{Figure}

Adjusting the quadrupole parameter leads to distinct variations in temperature fluctuations and trends in temperature change with distance, as depicted in Figures 40, 41, and 42. Specifically, we observe that a positive quadrupole generally results in a steeper temperature gradient, while a negative quadrupole tends to flatten the temperature profile. These observations suggest that the quadrupole parameter plays a significant role in shaping the temperature distribution within the system.

\begin{Figure}
 \centering
 \advance\leftskip-2cm
 \advance\rightskip-2cm
 \includegraphics[width=8cm, height=7cm]{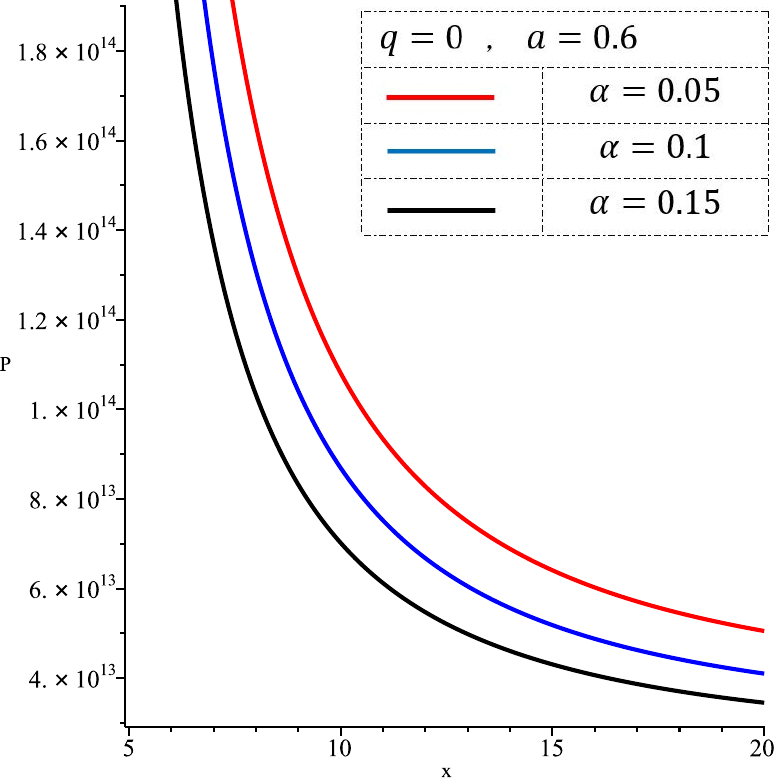}
 \captionof{figure}{Pressure for \(\alpha=0.01\) , \(\alpha=0.1\) and \(\alpha=0.5\) where \(q=0\).}
\end{Figure}

\begin{Figure}
 \centering
 \advance\leftskip-2cm
 \advance\rightskip-2cm
 \includegraphics[width=8cm, height=7cm]{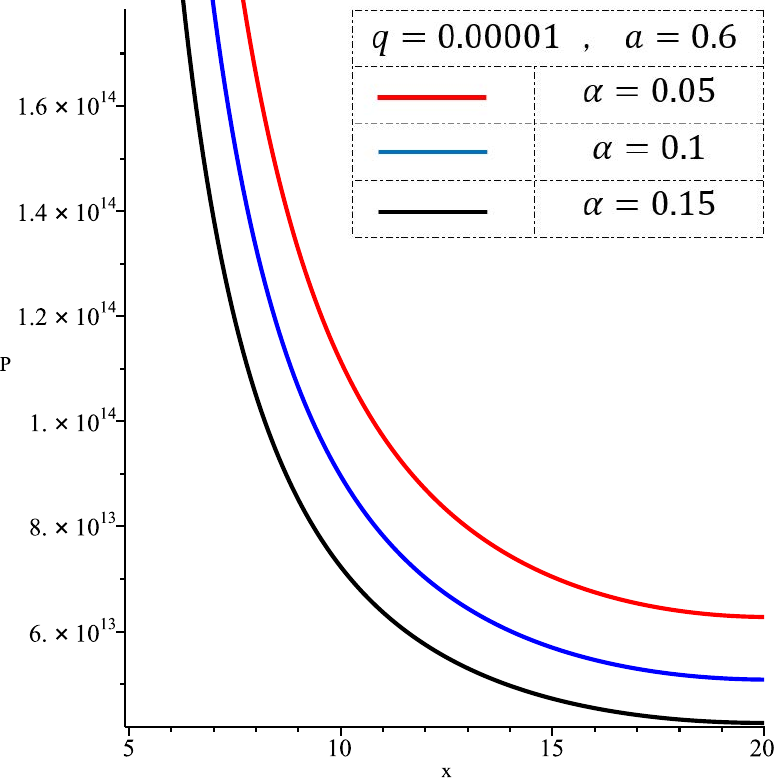}
 \captionof{figure}{Pressure for \(\alpha=0.01\) , \(\alpha=0.1\) and \(\alpha=0.5\) where \(q=0.00001\).}
\end{Figure}

\begin{Figure}
 \centering
 \advance\leftskip-2cm
 \advance\rightskip-2cm
 \includegraphics[width=8cm, height=7cm]{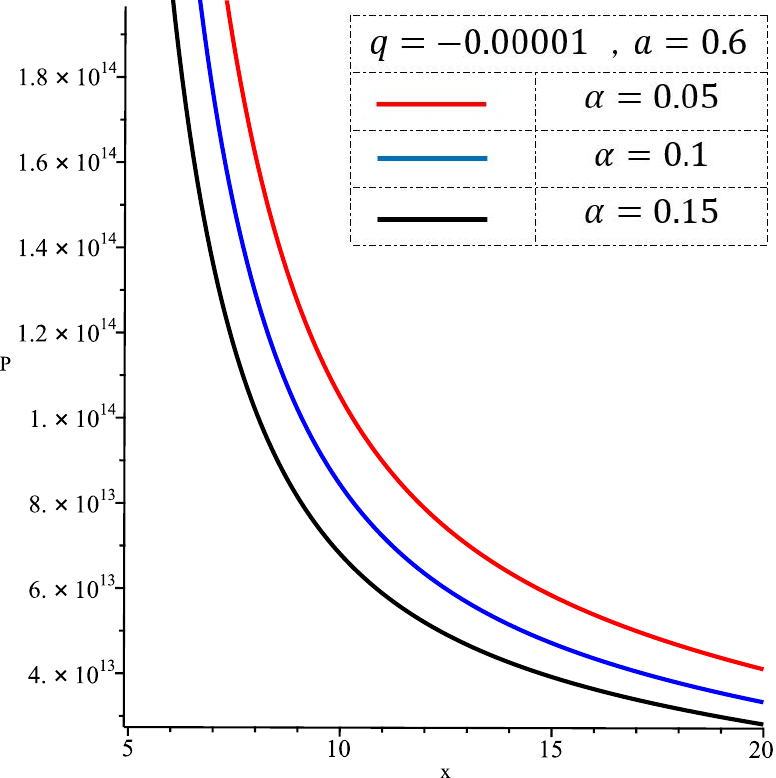}
 \captionof{figure}{Pressure for \(\alpha=0.01\) , \(\alpha=0.1\) and \(\alpha=0.5\) where \(q=-0.00001\).}
\end{Figure}

Analysis of the P-x diagrams reveals that increasing the \(\alpha\)\ parameter generally results in higher pressure values at certain distances compared to other states. This suggests a direct correlation between \(\alpha\)\ and pressure within the system. Further investigation into this relationship could provide valuable insights into the mechanisms by which \(\alpha\)\ influences pressure distribution.

\begin{Figure}
 \centering
 \advance\leftskip-2cm
 \advance\rightskip-2cm
 \includegraphics[width=8cm, height=6.9cm]{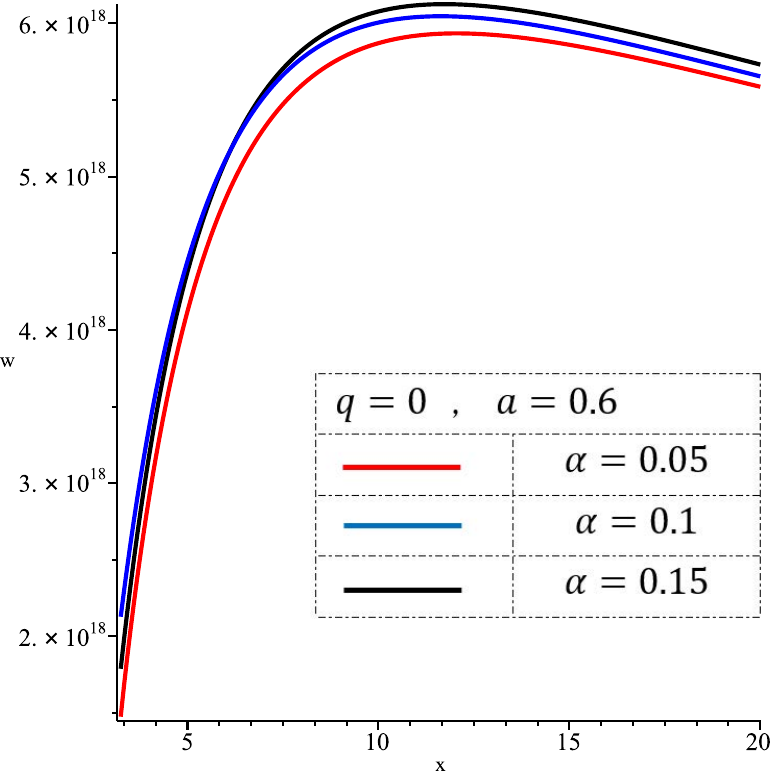}
 \captionof{figure}{viscous stress for \(\alpha=0.01\) , \(\alpha=0.1\) and \(\alpha=0.5\) where \(q=0\).}
\end{Figure}

\begin{Figure}
 \centering
 \advance\leftskip-2cm
 \advance\rightskip-2cm
 \includegraphics[width=8cm, height=6.9cm]{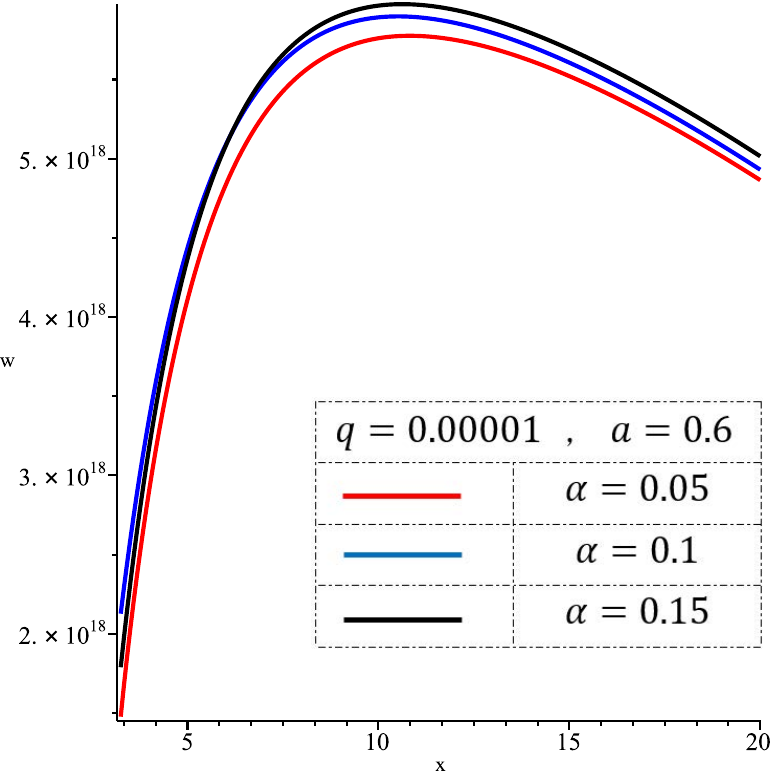}
 \captionof{figure}{viscous stress for \(\alpha=0.01\) , \(\alpha=0.1\) and \(\alpha=0.5\) where \(q=0.00001\).}
\end{Figure}

\begin{Figure}
 \centering
 \advance\leftskip-2cm
 \advance\rightskip-2cm
 \includegraphics[width=8cm, height=6.9cm]{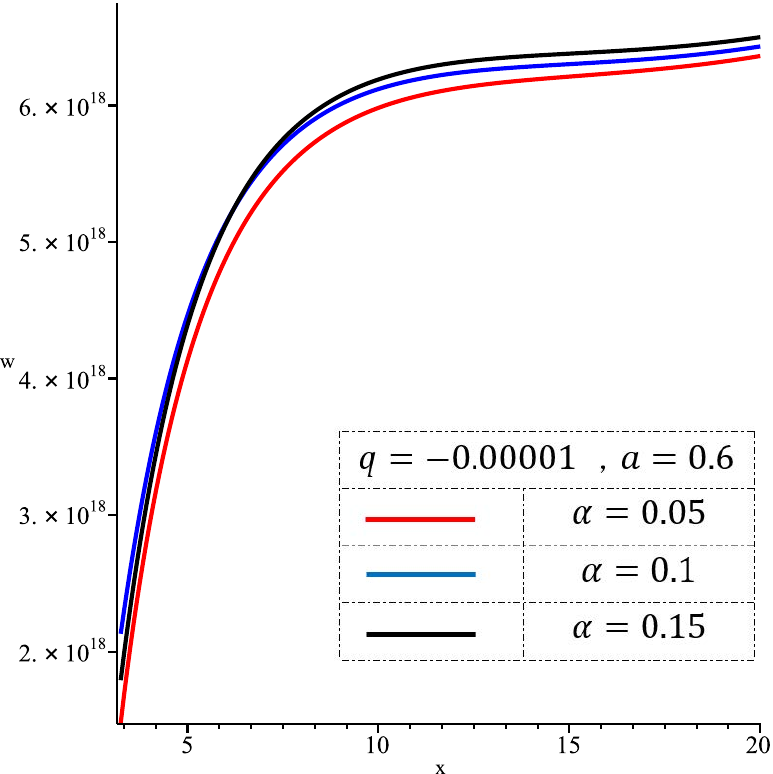}
 \captionof{figure}{viscous stress for \(\alpha=0.01\) , \(\alpha=0.1\) and \(\alpha=0.5\) where \(q=-0.00001\).}
\end{Figure}

In the w-x diagrams (Figures 46, 47, and 48), different quadrupoles have had significantly varied effects at fixed values of a and q. Each quadrupole alters the trend of the graph after the initial maximum point. Considering the observational data, the importance of utilizing each quadrupole with its respective value (positive, negative, or zero) becomes evident.

\section{Discussion and conclusion}\label{sec.discus and conclud}
In this article, we first examined the changes in the quadrupole moment at a fixed \(\alpha\) for both rotating and non-rotating states. Next, we varied alpha for quadrupoles with values of zero, positive, and negative, while keeping the rotation parameter at a positive value less than zero. Subsequently, we fixed alpha for the zero, positive, and negative quadrupoles and adjusted the rotation parameter. By plotting all relevant graphs, we analyzed their behaviors and effects.\\
\\
A detailed examination of the results reveals the following trends:\\
\\
1- Quadrupole Influence: The greater the deviation of the quadrupole moment (q) from zero, whether positive or negative, the more significant the deviation of the graphs from the base state.\\
\\
2- Specific Quantity Trends:\\
a) F-x graph: Increasing rotation (a) and viscosity (\(\alpha\)) leads to a higher peak in the flux-radial coordinate (F-x) graph. While different quadrupoles do not alter the maximum value, they do influence the shape of the graph over time.\\ 
b) \(\Sigma\)-x graph: An increase in rotation (a) and a decrease in viscosity (\(\alpha\)) result in an additional peak in the surface density-radial coordinate (\(\Sigma\)-x) graph. A positive quadrupole accelerates the growth of the graph, while a negative quadrupole exhibits slower growth.\\
c) h-x graph: Higher rotation (a) shifts the peak of the h-x graph closer to the horizon, but the peak value remains unchanged. Changes in \(\alpha\) and quadrupole have minimal effects on the h-x graph.\\
d) T-x graph: Increasing rotation (a) and viscosity (\(\alpha\)) elevate the average temperature in the temperature-radial coordinate (T-x) graph. Changes in quadrupole primarily affect the slope of the graph over time, with no noticeable impact at the initial stages.\\
e) P-x graph: As rotation (a) and viscosity (\(\alpha\)) increase, pressure (P) also rises, and the difference between minimum and maximum pressure values expands. A positive quadrupole accelerates pressure growth, while a negative quadrupole leads to slower pressure growth.\\
f) w-x graph: Increasing rotation (a) and viscosity (\(\alpha\)) lead to greater changes in viscosity per quadrupole in the viscosity-radial coordinate (w-x) graph. The slope of the w-x graph decreases after the peak for a positive quadrupole, whereas it remains at the maximum value for a negative quadrupole.\\
\\
This result can be understood by considering the effect of the quadrupole moment on the geometry of the black hole's spacetime. The quadrupole moment introduces an oblateness or prolateness to the shape of the event horizon, which in turn affects the motion of particles and radiation near the black hole. In particular, the quadrupole moment can affect the efficiency of energy extraction from the black hole's spin, a key factor in powering the emission from accretion disks.
\newpage
\bibliographystyle{unsrt}
\bibliography{mybibfile}

\end{multicols}

\newpage
\appendix
\section{APPENDIX}\label{sec.APPEN A}

\setcounter{equation}{0}
\renewcommand{\theequation}{A-\arabic{equation}}

According to (1) we can calculate $g_{\alpha \beta}$ and $\Gamma$ :
\begin{equation}
g_{tt}=-\frac{A}{B} e^{2U}.
\end{equation}
\begin{equation}
g_{t\phi}= \frac{2\omega A}{B} e^{2U}.
\end{equation}
\begin{equation}
g_{\phi \phi}= -\frac{\omega^2 A}{B} e^{2U}+\frac{B e^{-2U} (x^2-1) (1-y^2)}{A}.
\end{equation}

According to Metric $g_{\alpha \beta}$ and \cite{harko2011thin}
\begin{equation}
\Omega=\frac{-(g_{t\phi})_{,r}+\sqrt{(-(g_{t\phi})_{,r})^2-(g_{\phi\phi})_{,r}(g_{tt})_{,r}}}{g_({\phi \phi})_{,r}}.
\end{equation}
\begin{equation}
E=\frac{g_{tt}+\Omega g_{t\phi}}{\sqrt{-g_{tt}-2\Omega g_{t\phi}-\Omega^2 g_{\phi\phi}}}.
\end{equation}
\begin{equation}
L=\frac{\Omega g_{\phi\phi}+g_{t\phi}}{\sqrt{-g_{tt}-2\Omega g_{t\phi}-\Omega^2 g_{\phi\phi}}}.
\end{equation}

By putting equations (30), (31) and (32) in these relations, equations (33) to (35) are obtained.To accurately calculate these relationships in terms of q, the results will be very long, but it can be easily calculated using programming software such as Maple and Mathematica.

Now we want to calculate the Coefficients for Kerr metric in the quadrupole version. from \cite{thorne1974disk} we can use the equations and do following to find the results. to comparing both metric with q and without q, Coefficient of \(dt^2\) from \cite{page1974disk} and \cite{abdolrahimi2015properties}

\begin{equation}
-\frac{\mathbb{D}}{\mathbb{A}}+\frac{M^2 x^4 \mathbb{A} 4a_{*}^2}{M^2 x^{12} \mathbb{A}}=\frac{-\mathbb{D} x^8+4a_{*}^2}{x^8 \mathbb{A}} , -e^{2U} \frac{A}{B}.
\end{equation}
They are equal
\begin{equation}
\mathbb{A}=\frac{B}{x^8} , \mathbb{D}=\frac{1}{x^8}(A e^{2U}+4a_{*}^2).
\end{equation}
also 
\begin{equation}
\Omega=\frac{1}{M x^3 \mathbb{B}} \implies \mathbb{B}=\frac{1}{M x^3 \Omega}.
\end{equation}
and
\begin{equation}
(E-\Omega L)^2=(\frac{\sqrt{\mathbb{C}}}{\mathbb{B}})^2 \implies \mathbb{C}=\mathbb{B}^2 (E-\Omega L)^2.
\end{equation}
\begin{equation}
E=\frac{\mathbb{g}}{\sqrt{\mathbb{C}}} \implies \mathbb{g}=E \sqrt{\mathbb{C}}.
\end{equation}
\begin{equation}
L=\frac{M x \mathbb{F}}{\sqrt{\mathbb{C}}} \implies \mathbb{F}=\frac{L \sqrt{\mathbb{C}}}{M x}.
\end{equation}
\\

\end{document}